\documentclass[onecolumn,journal]{IEEEtran}
\IEEEoverridecommandlockouts
\usepackage{cite}
\usepackage{amsmath,amssymb,amsfonts}
\usepackage{algorithm}
\usepackage{algorithmic}
\usepackage{graphicx}
\usepackage{textcomp}
\usepackage{xcolor}
\usepackage{subfigure}
\usepackage[section]{placeins}
\newtheorem{remark}{\text{Remark}}
\newtheorem{lemma}{\text{Lemma}}

\begin{document}

\title{Power Measurement Based Channel Estimation for IRS-Enhanced Wireless Coverage}
\author{He Sun, \emph{Member, IEEE},
         Lipeng Zhu, \emph{Member, IEEE},
         Weidong Mei, \emph{Member, IEEE},
         and Rui Zhang, \emph{Fellow, IEEE}
\thanks{Part of this work has been presented at the IEEE Global Communications Conference (GLOBECOM), Kuala Lumpur, Malaysia \cite{sun}.} \\
\thanks{H. Sun and L. Zhu are with the Department of Electrical and Computer Engineering, National University of Singapore, Singapore 117583 (e-mail: sunele@nus.edu.sg, zhulp@nus.edu.sg). W. Mei is with National Key Laboratory of Wireless Communications, University of Electronic Science and Technology of China, China 611731 (e-mail: wmei@uestc.edu.cn). R. Zhang is with School of Science and Engineering, Shenzhen Research Institute of Big Data, The Chinese University of Hong Kong, Shenzhen, Guangdong 518172, China (e-mail: rzhang@cuhk.edu.cn). He is also with the Department of Electrical and Computer Engineering, National University of Singapore, Singapore 117583 (e-mail: elezhang@nus.edu.sg).}
}

\markboth{}%
{Shell \MakeLowercase{\textit{et al.}}: Bare Demo of IEEEtran.cls for IEEE Journals}

\maketitle

\begingroup
\allowdisplaybreaks
\vspace{-16pt}

\begin{abstract}
Intelligent reflecting surface (IRS) has been recognized as a transformative technology for enabling smart and reconfigurable radio environment cost-effectively by leveraging its controllable passive reflection. In this paper, we study an IRS-assisted coverage enhancement problem for a given region, aiming to optimize the passive reflection of the IRS for improving the average communication performance in the region by accounting for both deterministic and random channels in the environment. To this end, we first derive the closed-form expression of the average received signal power in terms of the deterministic base station (BS)-IRS-user cascaded channels over all user locations, and propose an IRS-aided coverage enhancement framework to facilitate the estimation of such deterministic channels for IRS passive reflection design. Specifically, to avoid the exorbitant overhead of estimating the cascaded channels at all possible user locations, a location selection method is first proposed to select only a set of typical user locations for channel estimation by exploiting the channel spatial correlation in the region. To estimate the deterministic cascaded channels at the selected user locations, conventional IRS channel estimation methods require additional pilot signals, which not only results in high system training overhead but also may not be compatible with the existing communication protocols. To overcome this issue, we further propose a single-layer neural network (NN)-enabled IRS channel estimation method in this paper, based on only the average received signal power measurements at each selected location corresponding to different IRS random training reflections, which can be offline implemented in current wireless systems. Based on the estimated channels, the IRS passive reflection is then optimized to maximize the average received signal power over the selected locations. Numerical results demonstrate that our proposed scheme can significantly improve the coverage performance of the target region and outperform the existing power-measurement-based IRS reflection designs.
\end{abstract}

\begin{IEEEkeywords}
Intelligent reflecting surface, channel estimation, passive reflection design, neural network, measurement location selection, spatial correlation.
\end{IEEEkeywords}

\section{Introduction}

Intelligent reflecting surface (IRS) has recently emerged as a promising candidate technology for the future six-generation (6G) wireless communication systems due to its capability of realizing smart and reconfigurable propagation environment in a cost-effective manner\cite{WQQTutorial,wu2023intelligent}. Specifically, an IRS consists of a large number of passive reflecting elements with independently tunable reflection coefficients, which can be jointly designed to alter the phase and/or amplitude of its incident signal to achieve high-performance passive beamforming/reflection for various purposes, such as signal boosting, interference suppression, target sensing, etc\cite{WQQTutorial,surveyapp,mei}.
\begin{figure}[t]
  \centering
  {\includegraphics[width=0.5\textwidth]{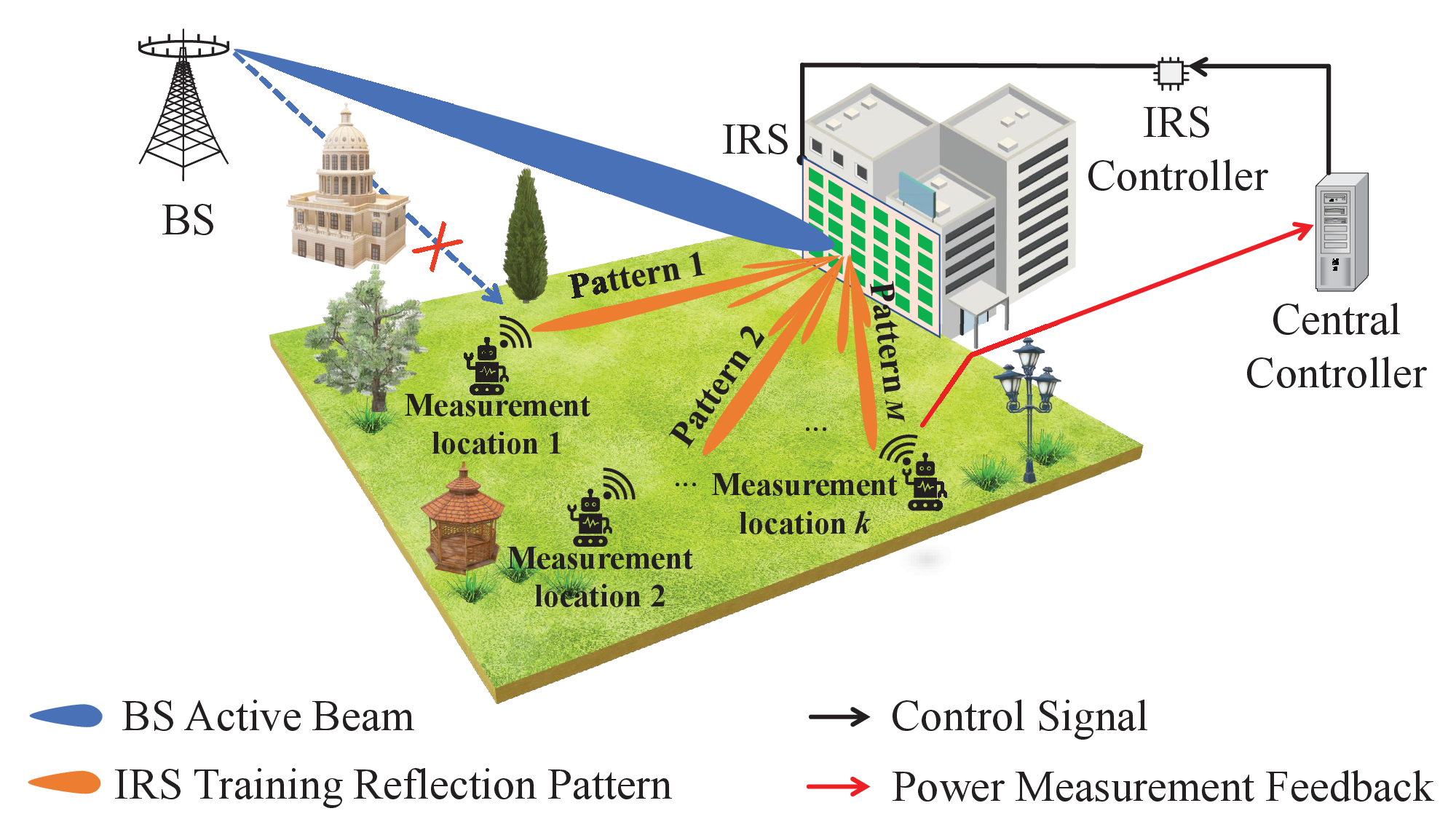}}
  \caption{IRS-enhanced communication coverage with power measurement feedback. }
\label{fig01001}
\vspace{-16pt}
\end{figure}


To achieve the above benefits, IRS passive beamforming or in general passive reflection needs to be properly designed, which has been extensively studied under various wireless system setups, such as multi-antenna/multiple-input multiple-output (MIMO)\cite{IRSMIMO,IRSMIMO2}, integrated sensing and communication (ISAC)\cite{JSAC22,10149442}, unmanned aerial vehicle (UAV)\cite{UAV,9777746}, physical layer security\cite{pls}, orthogonal frequency-division multiplexing (OFDM)\cite{IRSOFDM}, orthogonal time-frequency-space (OTFS) modulation\cite{IRSOTFS}, multi-cell network\cite{cells}, non-orthogonal multiple access (NOMA)\cite{IRSOMA}, and so on. In the existing literature, there are mainly two practical approaches for IRS passive beamforming/reflection design, which are based on channel estimation pilots and user signal power measurements, respectively. In the former approach, the cascaded base station (BS)-IRS-user/user-IRS-BS channels are first estimated based on the downlink/uplink pilots received at the users/BS with time-varying IRS training reflection patterns, and then the IRS reflection for data transmission is optimized based on the estimated IRS cascaded channels\cite{IRSCESurvey1,IRSCESurvey2,mei,DDNN}. Alternatively, the authors in \cite{YWJSAC} proposed to train a deep neural network (DNN) to directly learn the mapping from the received pilot signals to the optimal IRS reflection for data transmission. However, the above pilot-based designs require excessive training pilots for IRS channel estimation or DNN training, which not only increases the system training overhead but also may not be compatible with the existing wireless transmission protocols (e.g., for cellular network and WiFi) that cater to the user-BS direct channel (without IRS) estimation only. To efficiently integrate IRS into current wireless systems without the need of modifying their protocols, the latter approach designs IRS reflection for data transmission based on the received (pilot or data) signal power measurements at each user's receiver with time-varying IRS reflection patterns, as shown in Fig. \ref{fig01001}, which can be easily obtained in existing wireless systems.
For example, passive beam training for IRS-aided millimeter-wave (mmWave) systems\cite{You1,You2} and conditional sample mean (CSM)-based IRS reflection optimization for IRS-aided sub-6 GHz systems\cite{CSM,ACSM} have been proposed. In particular, it was shown in \cite{CSM,ACSM} that in the single-user case, the CSM method can achieve an IRS passive beamforming gain in the order of the number of IRS reflecting elements, which is identical to that under perfect channel state information (CSI)\cite{WQQTutorial}. However, the number of random IRS reflections required for CSM is usually very large (hundreds or even thousands) to obtain sufficient user power measurement samples for even the single-user case, which can result in high implementation overhead and large delay. The reason for such low efficiency mainly lies in the lack of IRS channel recovery from the users' power measurements in CSM.

Moreover, the above IRS passive reflection designs mainly aim to improve the communication performance for a given number of user locations based on their instantaneous CSI and thus need to be implemented over channel coherence time for adapting to the CSI variation in practice, which incurs high implementation complexity and large training overhead. As such, other works have proposed to exploit the long-term/statistical CSI in the IRS-aided communication system to facilitate the IRS passive reflection design \cite{IRSCESurvey2}, which varies much slower than the instantaneous CSI. Particularly, for the IRS-aided coverage performance optimization in a given region (see, e.g., Fig. \ref{fig01001}), as the IRS passive reflection design needs to cater to all possible user locations in the region, it is practically viable to exploit their long-term CSI to avoid the prohibitive overhead for instantaneous CSI estimation. For example, the authors in \cite{wenyan,outage} revealed that IRS can offer significant improvement in the coverage performance within a region. However, the IRS passive reflection in \cite{wenyan} was designed assuming known line-of-sight (LoS) channels from the IRS to any user location and the BS, while that in \cite{outage} considered a more general multi-path channel model but assumed perfect channel spatial correlation information between any two locations in the region, both of which are difficult to acquire in practice. Besides, the authors in\cite{weimultiirs,10439018} studied a multi-IRS deployment problem for coverage enhancement in a given region, but also under the assumption of known LoS channels to/from each IRS. Thus, there is still a lack of practically implementable IRS passive reflection design for coverage performance enhancement in a given region without any {\it a priori} known CSI.

To tackle the above problem, we propose in this paper a new IRS-aided communication coverage enhancement framework for a target region (see Fig. \ref{fig01001}) based on offline user power measurements. Our main contributions are summarized as follows.

\begin{itemize}
  \item First, under a general multi-path channel model accounting for both deterministic and random scatterers in the given region, we derive the long-term average received signal power at each user location in this region in terms of the deterministic BS-IRS-user cascaded channel, which can be estimated for IRS passive reflection design. However, estimating the cascaded channels for all user locations results in practically exorbitant overhead. To avoid such high overhead as well as the complexity of IRS passive reflection optimization, we propose a location selection algorithm which chooses only a set of typical locations for IRS channel estimation by exploiting the channel spatial correlation in the region. In particular, the entire region is divided into a set of subregions each with high channel spatial correlation to facilitate the power-measurement location selection.
  \item Furthermore, to acquire the cascaded channel information at the selected locations, we propose a new IRS cascaded channel estimation scheme based on the average received power measured at each selected location offline (e.g., by a mobile robot user terminal, as shown in Fig. \ref{fig01001}) with randomly generated IRS reflection patterns. Specifically, we reveal that for a given IRS reflection pattern, the received signal power at any user location can be equivalently modeled as the output of a simple single-layer neural network (NN), with its weights corresponding to the coefficients of the deterministic cascaded BS-IRS-user channel at that location. Inspired by this, we optimize the weights of the NN to minimize the mean squared error (MSE) between its output and each power measurement via the stochastic gradient descent method, thereby estimating the deterministic IRS cascaded channel at each location. As compared to the aforementioned CSM method for designing the IRS reflection which is also based on user terminals' power measurements, our proposed method further exploits them to recover the deterministic IRS channels, which thus enables more efficient IRS reflection design for coverage enhancement.
  \item Finally, the IRS passive reflection is optimized based on the estimated deterministic IRS channels at the selected locations to maximize their average received signal power, thereby enhancing the communication coverage in the whole region. Simulation results under the practical ray-tracing based propagation environment show that our proposed framework can yield much better communication coverage performance as compared to the existing power-measurement-based methods for IRS passive reflection design such as CSM.
\end{itemize}

The rest of this paper is organized as follows. Section II presents the system model. Section \ref{sec003} presents the problem formulation for coverage improvement and the proposed IRS-aided communication coverage enhancement framework. Section \ref{sec005} presents the proposed power-measurement location selection algorithm. Section \ref{sec004} presents the proposed single-layer NN for IRS channel estimation. Section \ref{sec007} presents numerical results to verify the efficacy of our proposed framework. Finally, Section \ref{sec008} concludes this paper.

\emph{Notations}: Scalars, vectors and matrices are denoted by lower/upper case, boldface lower case and boldface upper case letters, respectively. For any scalar/vector/matrix, $(\cdot)^*$, $(\cdot)^T$ and $(\cdot)^H$ respectively denote its conjugate, transpose and conjugate transpose. $\mathbb{C}^{n \times m}$ and $\mathbb{R}^{n \times m}$ denote the sets of $n \times m$ complex and real matrices, respectively. $\cap$ denotes the intersection of two sets. $\parallel\cdot\parallel$ denotes the Euclidean norm of a vector, and $\left\|\cdot\right\|_F$ denotes the Frobenius norm of a matrix. $\mid\cdot\mid$ denotes the cardinality of a set or the amplitude of a complex number. $\left\lceil\cdot\right\rceil$ denotes the smallest integer no lower than its argument. $\jmath=\sqrt{-1}$ denotes the imaginary unit. $\nabla_{\boldsymbol x}F \in \mathbb{R}^{n}$ denotes the gradient of a function $F$ with respect to its argument ${\boldsymbol x} \in \mathbb{R}^{n}$. $\rm Re(\cdot)$ and $\rm Im(\cdot)$ denote the real and imaginary parts of a complex vector/number, respectively. $\text{Tr}(\cdot)$ denotes the trace of a matrix. The distribution of a circularly symmetric complex Gaussian (CSCG) random variable with mean zero and covariance $\sigma^2$ is denoted by $\mathcal{CN}(0,\sigma^2)$.

\section{System Model }\label{sec002}

As shown in Fig. \ref{fig01001}, we consider a wireless network in a given subregion $\cal{D}$ served by a remote multi-antenna BS with its beam fixed (i.e., with a single antenna equivalently) to cover the region of interest (e.g., in a given sector of the serving BS). Due to the dense blockage (e.g., buildings), we assume that the direct links between the BS and locations in $\cal{D}$ are severely blocked.\footnote{The results in this paper can be readily extended to the case with the BS-user direct links considered, by following a similar system model as that in \cite{CSM,ACSM}.} To enhance the BS's coverage for users in this region, an IRS equipped with $N$ passive reflecting elements is deployed to assist in the wireless communications between the BS and users located in $\cal{D}$.

For ease of exposition, we divide the region $\cal{D}$ into $K$ squared grids of equal size $d_0 \times d_0$, which is sufficiently small such that the users at each grid are assumed to experience the same deterministic channel with the IRS due to static scatterers in the region, while they may also experience time-varying random channels with the IRS due to moving scatterers.

Let $\mathbf{\Theta}=\text{diag}(e^{\jmath{\theta}_1}, \cdots ,e^{\jmath{\theta}_N})$ denote the reflection matrix of the IRS, where ${\theta}_i$ denotes the phase shift of its $i$-th reflecting element, $1 \leq i \leq N$. Due to the hardware constraints, we consider that the phase shift of each IRS reflecting element can only take a finite number of discrete values, i.e.,
\begin{equation}\label{eqssec2001}
  {\theta}_n \in \Phi_\alpha \triangleq \{\omega, {2\omega},{3\omega}, \cdots ,2^\alpha{\omega}\}, \forall n,
\end{equation}
where $\alpha$ is the number of bits for controlling the discrete phase shift, and $\omega = \frac{2\pi}{2^\alpha}$.

Let ${\cal{K}} \triangleq \left\{1,2, \cdots, K\right\}$ denote the set of grids in the region $\cal D$. Let $\boldsymbol{h}_{BI} \in {\mathbb{C}}^{N \times 1}$ and $\boldsymbol{h}_{I{U_k}}^H \in {\mathbb{C}}^{1 \times N}$ denote the baseband equivalent channel from the BS to the IRS and that from the IRS to any location in grid $k$ (denoted as $U_k, k \in {\cal K}$), respectively. Accordingly, the effective channel from the BS to $U_k$ via the IRS is expressed as
\begin{equation}\label{eqs01001}
  g_k = {\boldsymbol{h}_{I{U_k}}^H}\mathbf{\Theta}\boldsymbol{h}_{BI}, \ k \in {\mathcal{K}}.
\end{equation}

In this paper, we consider the general multi-path IRS-$U_k$ and BS-IRS channels, each consisting of multiple deterministic paths due to the static/dominant scatterers, as well as multiple random paths due to other time-varying/non-dominant scatterers. Hence, the IRS-$U_k$ and BS-IRS channels can be respectively expressed as
\begin{equation}\label{01002}
\begin{split}
  {\boldsymbol{h}_{I{U_k}}^H} &= \sqrt{\frac{\varepsilon_{I{U_k}}}{1+\varepsilon_{I{U_k}}}}\boldsymbol{\overline{h}}_{I{U_k}}^H + \sqrt{\frac{1}{1+\varepsilon_{I{U_k}}}}\boldsymbol{{\widetilde{h}}}_{I{U_k}}^H,\\
  {\boldsymbol{h}_{B{I}}} &= \sqrt{\frac{\varepsilon_{BI}}{1+\varepsilon_{BI}}}\boldsymbol{\overline{h}}_{BI} + \sqrt{\frac{1}{1+\varepsilon_{BI}}}\boldsymbol{{\widetilde{h}}}_{BI},
\end{split}
\end{equation}
where ${\boldsymbol{\overline{h}}}_{I{U_k}}^H$ and ${\boldsymbol{\overline{h}}}_{BI}$ denote the superpositions of all deterministic paths for the IRS-$U_k$ and BS-IRS channels, respectively; $\boldsymbol{\widetilde{h}}_{I{U_k}}^H$ and $\boldsymbol{\widetilde{h}}_{BI}$ denote those of the other random paths for the IRS-$U_k$ and BS-IRS channels, respectively; and $\varepsilon_{I{U_k}}$ and $\varepsilon_{BI}$ denote the average power ratios of the deterministic paths to the random paths in the IRS-$U_k$ and BS-IRS channels, respectively.
In addition, the deterministic path components in (\ref{01002}), i.e., $\boldsymbol{\overline{h}}_{I{U_k}}^H$ and $\boldsymbol{\overline{h}}_{BI}$, follow the geometric channel model. Specifically, let $\boldsymbol{e}(\gamma,n)=[1, e^{-\jmath\pi\gamma}, e^{-\jmath2\pi\gamma}, \cdots ,e^{-\jmath{(n-1)}\pi\gamma}]^T$ denote the steering vector function of a uniform linear array (ULA) with $n$ elements and directional cosine $\gamma$. We assume that the IRS is equipped with a uniform planar array (UPA) parallel to the $y$-$z$ plane. Let $N_y$ and $N_z$ denote the numbers of reflecting elements along the axes $y$ and $z$, respectively. We define $\boldsymbol{u}_N(\vartheta,\varphi)= \boldsymbol{e}({\rm sin}(\vartheta){\rm sin}(\varphi), N_y)\otimes\boldsymbol{e}({\rm cos}(\vartheta), N_z)$ as the steering vector of a UPA with $\vartheta \in [0,\pi]$ and $\varphi\in [0,\pi]$ denoting the azimuth and elevation angles of arrival/departure (AoA/AoD) with respect to it. The deterministic channel components in (\ref{01002}) are thus expressed as
\begin{subequations}\label{eqs0204}
\begin{align}
  {\boldsymbol{\overline{h}}}_{I{U_k}}^H \ & =\sum_{l=1}^{L_{k,1}}\sqrt{\beta_{I{U_k}}}{e^{\frac{-\jmath2\pi {d_{l,I{U_k}}}}{\lambda}}}{\boldsymbol{u}_N(\vartheta_{l,k},\varphi_{l,k})},\label{Zg}\\
  {\boldsymbol{\overline{h}}}_{BI} \ & =\sum_{l=1}^{L_{2}}\sqrt{\beta_{BI}}{e^{\frac{-\jmath2\pi {d_{l,BI}}}{\lambda}}}{\boldsymbol{u}_N(\psi_{l},\phi_{l})},
\end{align}
\end{subequations}
where $\lambda$ denotes the signal wavelength, ${L_{k,1}}$ denotes the number of deterministic paths due to static/dominant scatterers between the IRS and $U_k$, with $d_{l,I{U_k}}$, $\vartheta_{l,k} \in [0,\pi]$ and $\varphi_{l,k}\in [0,\pi]$ denoting the length, the azimuth and elevation AoDs for the $l$-th path, respectively, ${L_{2}}$ denotes that between the BS and the IRS with $d_{l,BI}$, $\psi_{l} \in [0,\pi]$ and $\phi_{l} \in [0,\pi]$ denoting the length, the azimuth and elevation AoAs for the $l$-th path, respectively, $\beta_{I{U_k}}$ and $\beta_{BI}$ denote the large-scale power gains of the IRS-$U_k$ channel and the BS-IRS channel, respectively. On the other hand, for the random components in (\ref{01002}), i.e., $\boldsymbol{{\widetilde{h}}}_{I{U_k}}^H$ and $\boldsymbol{{\widetilde{h}}}_{BI}$, we assume that the number of time-varying/non-dominant scatterers is sufficiently large, such that their effect can be approximately modeled as complex Gaussian distribution based on the central limit theorem. Hence, we assume
\begin{subequations}
\begin{align}
  \boldsymbol{\widetilde{h}}_{I{U_k}} & \sim \mathcal{CN}(\textbf{0},{\beta_{I{U_k}}}\textbf{\emph{I}}_N),\label{Zh}\\
  \boldsymbol{\widetilde{h}}_{BI} & \sim \mathcal{CN}(\textbf{0},{\beta_{BI}}\textbf{\emph{I}}_N).
\label{e005}
\end{align}
\end{subequations}
Next, let ${{\boldsymbol{v}}}^H=\left[e^{\jmath{\theta}_1}, \cdots ,e^{\jmath{\theta}_N}\right]$ denote the passive reflection vector of the IRS, and ${\boldsymbol{h}}_k={\text{diag}}({\boldsymbol{h}_{I{U_k}}^H}){\boldsymbol{h}_{BI}}$ denote the cascaded BS-IRS-$U_k$ channel. As such, the channel in (\ref{eqs01001}) can be simplified as
\begin{equation}\label{eqs0205}
  g_k = {\boldsymbol{v}}^H{\boldsymbol{h}}_k, \ k \in {\mathcal{K}}.
\end{equation}
Let $s \in \mathbb{C}$ denote the transmit symbol at the BS with $\mathbb{E}[\left\vert s \right\vert^2]=1$. The received signal at $U_k$ is thus given by
\begin{equation}\label{eqs1006}
  y_{k} = \sqrt{P}{g_{k}}{s} + \emph{n}_k, \ k \in {\mathcal{K}},
\end{equation}
where $P$ denotes the transmit power at the BS, and $n_k\sim\mathcal{CN}(0,\sigma^2)$ denotes the complex additive white Gaussian noise (AWGN) at $U_k$ with power ${\sigma}^2$.

\section{Problem Formulation and Proposed Coverage Enhancement Framework }\label{sec003}

\subsection{Problem Formulation for Coverage Enhancement }
In this paper, we aim to improve the coverage performance over all the $K$ grids within the considered region $\cal D$. For each grid $k$, we focus on the long-term received signal power for users in it, i.e., $\mathbb{E}\left[|y_k|^2\right]$, where the expectation is taken over both the random channel components in (\ref{e005}) as well as the AWGN at the users' receiver. As such, we aim to optimize the IRS passive reflection to maximize the average of long-term received signal powers over all $K$ grids in $\cal D$. The associated optimization problem is formulated as
\begin{subequations}\label{eqs4001}
\begin{align}
  \text{(P1):} &\max_{\boldsymbol{v}} \ \frac{1}{K}\sum_{{{k \in \mathcal{K}}}} \ \mathbb{E}\left[|y_k|^2\right]\label{ZaP2}\\
& \text{s.t.} \ \ \ {\theta}_i\in \Phi_\alpha, i=1,2,\cdots,N. \label{Zb}
\end{align}
\end{subequations}

To solve (P1), we first derive the expectation in (\ref{ZaP2}). Based on (\ref{eqs1006}), the received signal power at $U_k$ is given by
\begin{equation}\label{eqs03002}
  \left|y_k\right|^2 = \left|\sqrt{P}g_{k}s + n_k\right|^2.
\end{equation}
By taking the expectation at both sides of (\ref{eqs03002}), we have
\begin{equation}\label{eqs03003}
  \mathbb{E}\left[\left|y_k\right|^2\right] = P\mathbb{E}\left[\left|g_k\right|^2\right] + \sigma^2,
\end{equation}
where we have utilized the fact that $\mathbb{E}\left[g_kn_k\right]=0$. To derive the expectation of $|g_k|^2$, we substitute (\ref{01002}) and (\ref{eqs0205}) into (\ref{eqs03003}) and obtain
\begin{equation}\label{eqs03004}
\begin{split}
  \mathbb{E}\left[\left|g_k\right|^2\right] & = {a_{1,k}}\left|{\boldsymbol{v}}^H{{{\boldsymbol{{h}}}}}^{'}_k\right|^2 + {a_{2,k}}\left\|\boldsymbol{\overline{h}}_{I{U_k}}^H\right\|^2 \\
    & \ \ \ \ + {a_{3,k}}\left\|\boldsymbol{\overline{h}}_{BI}^{}\right\|^2 + {a_{4,k}},
\end{split}
\end{equation}
where ${{{\boldsymbol{h}}}}^{'}_k\triangleq{\text{diag}}({\boldsymbol{\overline{h}}_{I{U_k}}^H}){\boldsymbol{\overline{h}}_{BI}}$ is defined as the deterministic cascaded channel at the $k$-th grid and
\begin{equation}\label{eqs03005}
\begin{split}
    a_{1,k} & = {\frac{\varepsilon_{I{U_k}}\varepsilon_{BI}}{(1+\varepsilon_{I{U_k}})(1+\varepsilon_{BI})}},\\
    a_{2,k} & = {\frac{\varepsilon_{I{U_k}}\beta_{BI}}{(1+\varepsilon_{I{U_k}})(1+\varepsilon_{BI})}},\\
    a_{3,k} & = {\frac{\varepsilon_{BI}\beta_{I{U_k}}}{(1+\varepsilon_{I{U_k}})(1+\varepsilon_{BI})}},\\
    a_{4,k} & = {\frac{N\beta_{BI}\beta_{I{U_k}}}{(1+\varepsilon_{I{U_k}})(1+\varepsilon_{BI})}}.
\end{split}
\end{equation}
The details of deriving (\ref{eqs03004}) are given in Appendix. Based on the above, for any given IRS reflection $\boldsymbol{v}$, we have
\begin{equation}\label{eqs0307}
\begin{split}
  \mathbb{E}\left[|y_k|^2\right] & = \left|{\boldsymbol{v}}^H{\boldsymbol{\overline{h}}}_k\right|^2 + C_{k},
\end{split}
\end{equation}
where ${\boldsymbol{\overline{h}}}_k=\sqrt{Pa_{1,k}}{{{\boldsymbol{h}}}}^{'}_k$ denotes the equivalent deterministic cascaded channel scaled by the factor $\sqrt{Pa_{1,k}}$, and $C_{k} = P{a_{2,k}}\left\|\boldsymbol{\overline{h}}_{I{U_k}}^H\right\|^2 + P{a_{3,k}}\left\|\boldsymbol{\overline{h}}_{BI}^{}\right\|^2 + P{a_{4,k}} + \sigma^2$ is a constant regardless of the IRS passive reflection.

By noting that the IRS reflection vector $\boldsymbol{v}$ only affects the  first term of (\ref{eqs0307}), problem (P1) can be simplified as
\begin{subequations}\label{eqs0308}
\begin{align}
  \text{(P2):} &\max_{\boldsymbol{v}} \ \frac{1}{K}\sum_{{{k \in \mathcal{K}}}} \ \boldsymbol{v}^H\boldsymbol{{G}}_{k}\boldsymbol{v} \\
& \text{s.t.} \ \ \ {\theta}_i\in \Phi_\alpha, i=1, 2, \cdots ,N, \label{p2c}
\end{align}
\end{subequations}
where $\boldsymbol{{G}}_k=\boldsymbol{\overline{h}}_{k}\boldsymbol{\overline{h}}_{k}^H$ denotes the covariance matrix of the deterministic cascaded channel $\boldsymbol{\overline{h}}_k$. Problem (P2) is a discrete optimization problem and can be solved by a variety of methods (see e.g., \cite{WdIRS,OptYG}), given that the perfect channel knowledge for the $K$ grids (i.e., $\boldsymbol{\overline{h}}_k$ or $\boldsymbol{{G}}_k$, $k \in {\mathcal{K}}$) is available.

However, the above channel knowledge is difficult to acquire in practice due to the following two reasons. First, solving (P2) requires the perfect CSI for all the $K$ grids, which is prohibitive to be estimated in practice if the size of the region $\cal D$ is large, as this results in an excessively large number of grids $K$ each with a given size.
Second, it is generally difficult to estimate the IRS's cascaded channel for each grid due to its passive reflection nature\cite{WQQTutorial}. Note that even though $\boldsymbol{\overline{h}}_k$, $k \in \cal{K}$, is only the deterministic cascaded channel instead of the exact one (i.e., $\boldsymbol{h}_k, k \in \cal{K}$), its estimation still requires extra pilot transmission (e.g., by first estimating the exact channels and then taking their average to eliminate the effect of random channel components).

\begin{figure*}[t]
  \centering
  {\includegraphics[width=0.7839863\textwidth]{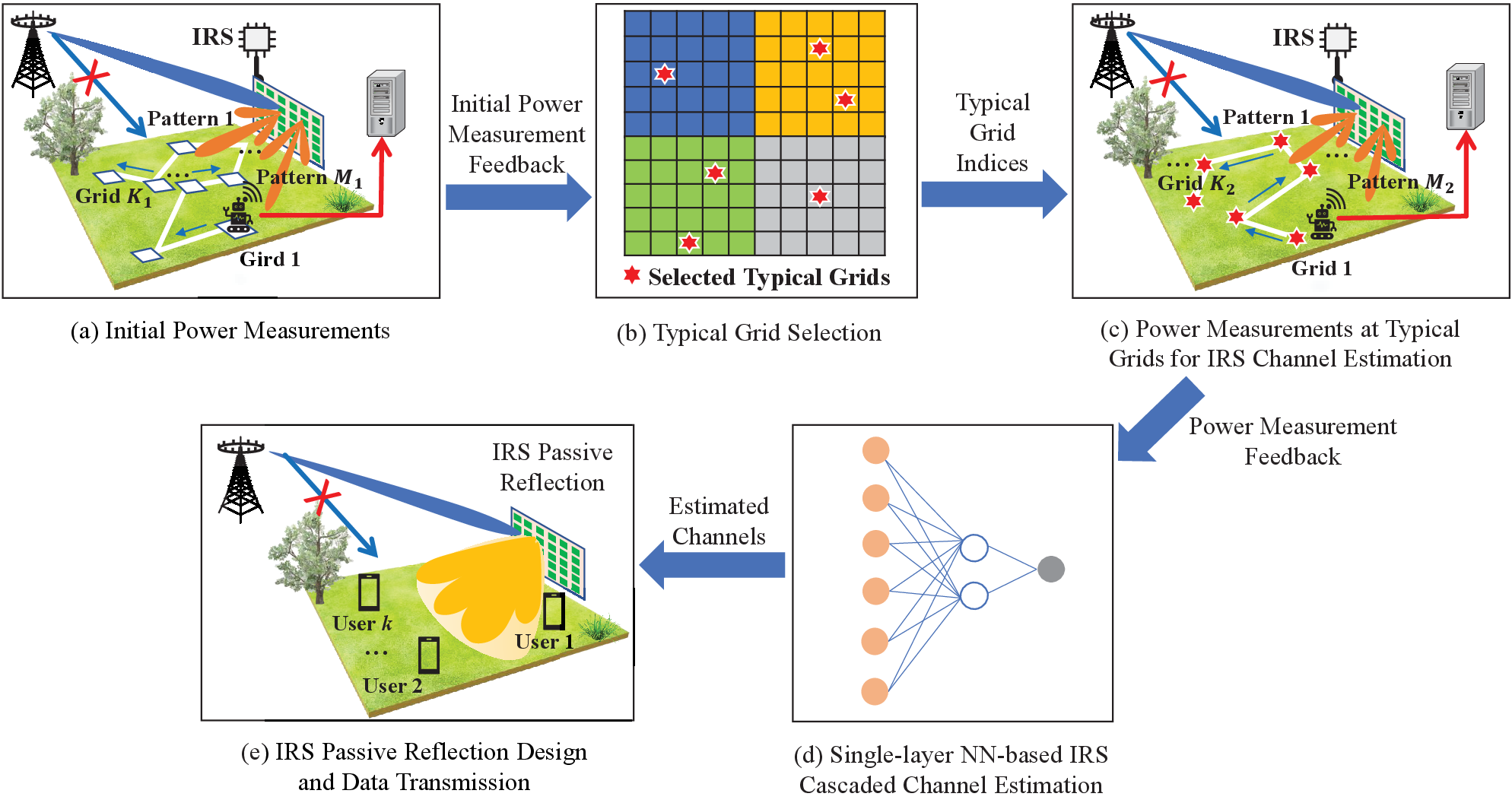}}
  \vspace{-10pt}\caption{Diagram of the proposed communication coverage enhancement framework. }
\label{nfig2}
\vspace{-10pt}
\end{figure*}

\subsection{Proposed Coverage Enhancement Framework }

To tackle the above challenges in solving (P2), we propose in this paper a new IRS-aided coverage enhancement framework, with its diagram shown in Fig. \ref{nfig2}. First, to dispense with the channel estimation at all $K$ grids, we exploit the channel spatial correlation among different grids in $\cal D$ to select only a set of representative grids for channel estimation. Notably, in current wireless systems (e.g., cellular and WiFi), user terminals are able to measure their received signal power, referred to as the reference signal received power (RSRP)\cite{3GPPTS38}\footnote{In practice, the RSRP measures the average received signal power over multiple frequency subcarriers. For ease of exposition, we assume in this paper narrowband transmission, while the proposed framework is extendable to wideband transmission over frequency-selective channels, which will be addressed in future work.}. By leveraging this capability, we propose to first characterize the channel spatial correlation in the region ${\cal D}$ by conducting initial power measurements at some grids\footnote{It is worth noting that since all locations in each grid $k$, $k \in \cal{K}$, are assumed to have the same deterministic IRS cascaded channel with the BS, the power measurement can be taken at any location in grid $k$. As such, we interchangeably use location and grid in the sequel of this paper without ambiguity.} in $\cal D$ under a small number of random IRS passive reflections (without IRS channel estimation), as shown in Fig. \ref{nfig2}(a). A central controller (e.g., the BS or another dedicated unit) collects such initial power measurement results via dedicated feedback links to characterize the channel spatial correlation and thereby selects a set of typical grids for estimating the deterministic IRS cascaded channels, as shown in Fig. \ref{nfig2}(b). Next, the indices of the selected typical grids are reported to the mobile terminals, which move to these grids for the subsequent channel estimation. The details of the above power-measurement location selection will be presented in Section \ref{sec005}.

Then, instead of using additional pilot transmission for estimating the deterministic IRS cascaded channel at each typical grid, we propose a power-measurement-based channel estimation method by leveraging a single-layer NN formulation, which only requires offline power measurements at the typical grids under a set of random IRS passive reflections as shown in Fig. \ref{nfig2}(c). After the power measurements at the typical grids, the central controller collects the measurement results from the mobile terminals to recover the deterministic IRS cascaded channels at these grids using the single-layer NN, as shown in Fig. \ref{nfig2}(d). The details of the IRS channel estimation will be provided in Section \ref{sec004}.

Finally, based on the channel estimates, the central controller optimizes the IRS passive reflection for coverage enhancement, by maximizing the average received signal power over the typical grids, i.e., replacing ${\cal K}$ and $K$ in (P2) with the set of typical grids and its cardinality, respectively. Such an optimized IRS passive reflection will be adopted for real-time data transmission (as shown in Fig. \ref{nfig2}(e)) and updated only if the (slowly varying) deterministic IRS cascaded channels in the region change.

\section{Power Measurement Location Selection}\label{sec005}

In this section, we present how to select the typical grids for IRS channel estimation. According to Tobler's first law of geography\cite{Tobler}, two locations tend to be spatially correlated if they are sufficiently close to each other. As such, the cascaded channels from the BS to two closely located grids via the IRS (e.g., ${\boldsymbol{\overline{h}}}_i$ and ${\boldsymbol{\overline{h}}}_j$ for grids $i$ and $j$) are spatially correlated in general due to their common BS-IRS channel and surrounding scatterers (thus similar multi-path effects). Based on (\ref{eqs0307}), such channel correlation between ${\boldsymbol{\overline{h}}}_i$ and ${\boldsymbol{\overline{h}}}_j$ also results in correlation between average received signal powers at these two grids, i.e.,  $\mathbb{E}\left[|y_i|^2\right]$ and  $\mathbb{E}\left[|y_j|^2\right]$, for any given IRS passive reflection $\boldsymbol{v}$. Motivated by this, we can leverage the power measurements to infer the channel spatial correlation between any two grids within the region, as detailed next.
\vspace{-9pt}

\subsection{Power Measurement at Any Grid}\label{sec004A}

To facilitate the characterization of the spatial correlation among different grids in $\cal D$, the IRS applies randomly generated phase shifts of its reflecting elements (subject to (\ref{eqssec2001})) to reflect the BS's signals to different grids. Consider that the IRS generates $M_1$ reflection sets and let ${\boldsymbol{v}}_m$, $m \in {\cal{M}}_1\triangleq \left\{1,2, \cdots ,M_1\right\}$, denote the $m$-th reflection set. In the meanwhile, the mobile terminal at grid $k$ (if any) conducts initial power measurements of its received signal power corresponding to each IRS reflection set (see Fig. \ref{nfig2}(a)). For each reflection set, we assume that $Q$ transmitted symbols are used by the BS for the mobile terminal to calculate the average received signal power. Note that in practice, it usually holds that $Q \gg 1$ since the IRS's reflection switching rate is usually much lower than the BS's symbol rate. Let ${\cal K}_1$ denote the set of grids for initial power measurement, with its cardinality $\left|{\cal K}_1 \right| = K_1 \ll K$. In this paper, we assume that these $K_1$ grids are randomly generated from all $K$ grids. The power measurement at grid $k, k \in {\cal K}_1,$ under the IRS's $m$-th reflection set is given by
\begin{equation}\label{eqs}
{\overline p}_k(\boldsymbol{v}_m) = \frac{1}{Q}\sum_{q=1}^Q{\left|{{\boldsymbol{v}}_m^H{\boldsymbol{h}}_k}{s} + \emph{n}_k(q)\right|^2}, \ k \in {\mathcal{K}_1}, \ m \in {\cal{M}}_1,
\end{equation}
where $\emph{n}_k(q)\sim\mathcal{CN}(0,\sigma^2)$ denotes the $q$-th sampled AWGN. Let $\boldsymbol{p}_k =\left[{\overline p}_k\left(\boldsymbol{v}_1\right), {\overline p}_k\left(\boldsymbol{v}_2\right), \cdots,  {\overline p}_k\left(\boldsymbol{v}_{M_1}\right)\right]$ denote the collection of initial power measurements at grid $k$ under the $M_1$ reflection sets of the IRS. After the above initial power measurements, the user terminal at grid $k$ reports $\boldsymbol{p}_k$ to the central controller for channel spatial correlation characterization, based on which the typical grids are selected for channel estimation.
\vspace{-9pt}

\subsection{Channel Spatial Correlation Characterization }\label{sec401}

By following the channel spatial prediction techniques \cite{Kriging}, we model the considered region as a stationary random field with the same mean power over its different grids. For any random field, the variogram characterizes the variance of the difference between the power measurements at any of its two grids and thus can be used to infer the spatial correlation. However, deriving such variogram requires power measurements at all grids in the region, which is practically prohibitive to implement. As such, we aim to estimate an empirical variogram function via curve fitting based only on the power measurements at the grids in ${\cal K}_1$. To this end, we first calculate the difference between the power measurements at any two grids $i$ and $j$, $i, j \in {\cal K}_1$, which is given by
\begin{equation}\label{eqs04002}
  V\left(i ,j\right) = {\left\|{\boldsymbol{p}}_{i}-{\boldsymbol{p}}_{j}\right\|}^2, \ i, j \in {\cal{K}}_1.
\end{equation}
It is worth noting that in (\ref{eqs04002}), to mitigate the effects of each specific IRS reflection set on the channel spatial correlation characterization, we adopt the power measurements under all of the $M_1$ random IRS reflection sets to calculate the power difference. It should also be mentioned that since the initial power measurements only aim to capture the channel spatial correlation in the region $\cal D$, which are coarse-grained as compared to those for IRS channel estimation as will be presented in Section \ref{sec004B}, the required number of IRS reflection sets (i.e., $M_1$) can be much smaller than that for IRS channel estimation.

Let $ \ell\left(i ,j\right)$ denote the distance between the centers of grids $i$ and $j$, which is referred to as \emph{lag distance}. Define $\Omega_d=\left\{\left(i ,j\right) \ | \ell\left(i ,j\right) = d, i,j \in {{\cal K}_1} \right\}$ as the set of pairs of grids with a lag distance of $d$. Based on (\ref{eqs04002}), the empirical variogram function can be obtained as\cite{worboysgis}
\begin{equation}\label{eqs04003}
  \gamma\left(d\right)=\frac{1}{\left|\Omega_d \right|}\sum_{\left(i ,j\right) \in \Omega_d}{V}\left(i ,j\right).
\end{equation}
However, due to the finite number of grids in ${\cal K}_1$, we can only calculate the empirical variogram for certain values of lag distances based on (\ref{eqs04003}) with $\left|\Omega_d \right|>0$. To estimate the true variogram value at any lag distance, a variety of empirical models can be used to fit the calculated variogram values, e.g., the exponential variogram model, Gaussian variogram model or spherical variogram model\cite{chi1999}.

As shown in Fig. \ref{fig04001}, as the lag distance becomes sufficiently large, the variogram will converge to a constant value referred to as \emph{sill}, indicating that the power measurements at two distant grids are approximately independent at this lag distance. The lag distance at which the variogram attains $95\%$ of the sill is defined as the \emph{correlated range}, denoted as $c^\star$, as illustrated in Fig. \ref{fig04001}. In general, any two locations within the correlated range are considered as highly correlated\footnote{If any two locations in a stationary random field have no spatial correlation, the variogram should be equal to a constant for any non-zero lag distance, and we have $c^\star=0$.}.

\begin{figure}[t]
\begin{center}
\hspace*{0.01cm}
\centering
\includegraphics[scale=0.43956]{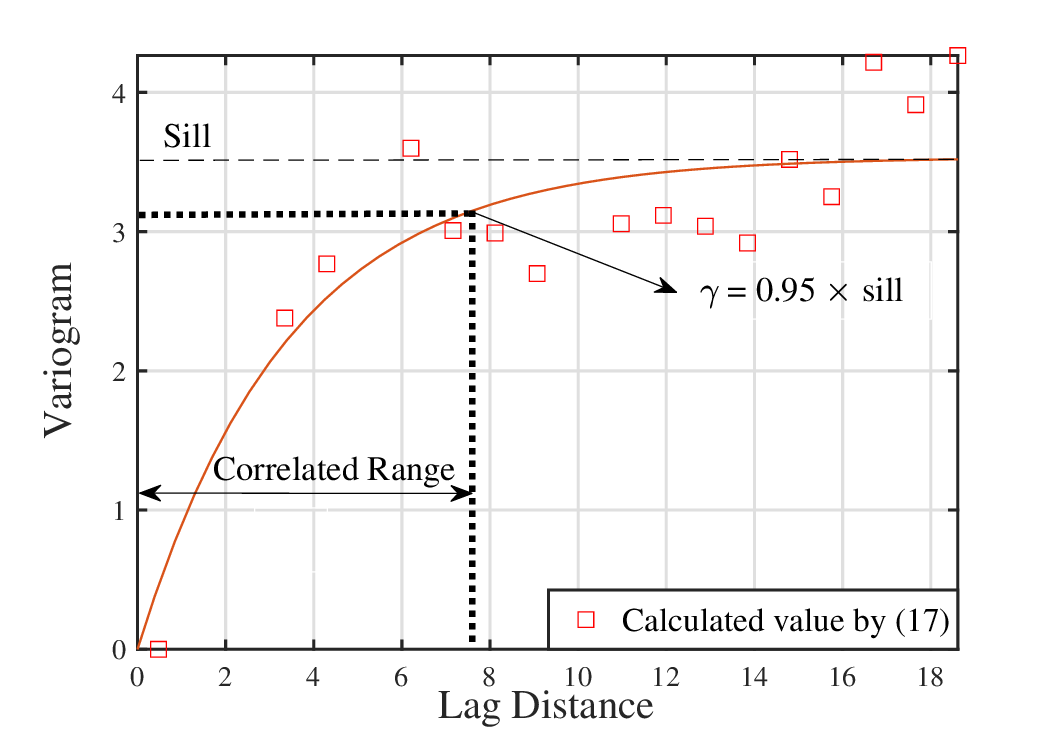} 
\caption{Illustration of the variogram function. }
\label{fig04001}\end{center}
\vspace{-9pt}
\end{figure}

\begin{figure}[t]
\begin{center}
\includegraphics[ scale = 0.3369 ]{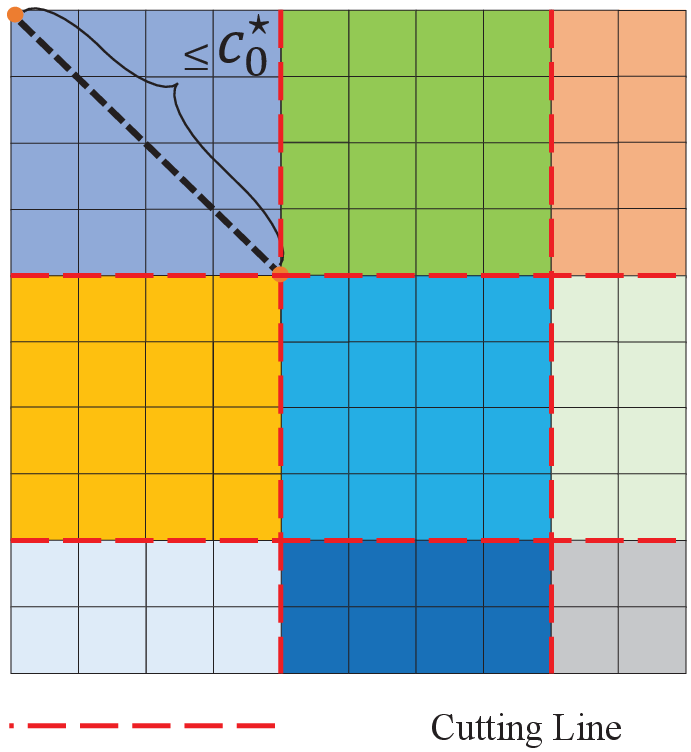} 
\caption{Illustration of the region splitting. }
\label{fig04002}
\vspace{-16pt}
\end{center}
\end{figure}

Based on the estimated empirical variogram function, we can estimate the correlated range of ${\cal D}$, denoted as $c_0^\star$, and thereby split it into multiple subregions with high channel correlation.
Let $\cal A$ denote the set of grids in any highly-correlated subregion, for which it must hold that
\begin{equation}\label{eqs00503}
  \ell\left(i ,j\right) \leq c^\star_0, \ \ \forall i, j \in \cal A.
\end{equation}
To ensure seamless coverage of the entire region, we consider in this paper that each subregion $\cal A$ has a square shape. As such, to satisfy (\ref{eqs00503}), the length of the diagonal of each square subregion should be no larger than $c^\star_0$, i.e., its side length should be no larger than $c^\star_0/\sqrt{2}$, as illustrated in Fig. \ref{fig04002}. Let $D_1$ and $D_2$ denote the side length of the region $\cal D$ in its two dimensions, respectively. To achieve the above region splitting, we can split these two dimensions into $\rho_1 = \left\lceil{ \frac{D_1}{{c^\star_{0}}/\sqrt{2}}}\right\rceil$ and $\rho_2 = \left\lceil{ \frac{D_2}{{c^\star_{0}}/\sqrt{2}}}\right\rceil$ segments, respectively, as shown in Fig. \ref{fig04002}. The total number of highly correlated subregions in $\cal D$ is given by $\rho_1 \rho_2$. Since the side length of each grid is $d_0$, the number of grids within each subregion is given by
\begin{equation}\label{eqs05004}
S_{\cal A}  = \left\lceil{ \frac{D_1}{{\rho_1 d_0}}}\right\rceil \times \left\lceil{ \frac{D_2}{{\rho_2 d_0}}}\right\rceil.
\end{equation}
It is worth noting that $D_1/(\rho_1 d_0)$ or $D_2/(\rho_2 d_0)$ may not be an integer. In this case, the subregions near the boundary of $\cal D$ may take a rectangular shape and have a smaller size than that of the other subregions. In the illustrative example of Fig. \ref{fig04002}, the region $\cal D$ is split into $9$ subregions, with $\rho_1=\rho_2=3$, and each subregion comprises at most $S_{\cal A}=16$ grids.

\subsection{Typical Grid Selection }

In a practical environment with diverse propagation conditions, the spatial correlation may vary across different local subregions in $\cal D$. Thus, the above region splitting based on global spatial correlation characterization may not result in the desired high correlation in each local subregion. In light of this, we further refine the estimation of correlated range for each subregion and select the typical grids in it accordingly. Let $T=\rho_1\rho_2$ denote the number of subregions after the region splitting in Section \ref{sec401}. Denote by ${\cal A}_{t}$ and $c^\star_{t}$ the set of grids in the $t$-th subregion and its correlated range, respectively. The correlated range $c^\star_{t}$ of the $t$-th subregion can be calculated via (\ref{eqs04003}) by replacing ${\cal K}_1$ therein with ${\cal A}_{t} \cap {\cal K}_1$.\footnote{For convenience, we assume ${\cal A}_{t} \cap {\cal K}_1 \ne \emptyset$. Otherwise, we can simply set $c_t^\star$ as the maximum distance between any two grids in this subregion.} Generally, the subregion with a small correlated range $c^\star_{t}$ shows more diverse propagation conditions. Hence, more grids should be selected from this subregion to extract richer environment features. Based on this observation, we propose the following rule to determine the number of typical grids selected from each subregion, i.e.,
\begin{equation}\label{a01}
  \eta_t = \left\lceil{\varrho \frac{\left|{\cal A}_{t}\right|d_0^2}{{c^\star_{t}}^2}}\right\rceil, \ \ \ t=1,2,\cdots, T,
\end{equation}
where $\left|{\cal A}_{t}\right|d_0^2$ denotes the size of the $t$-th subregion, and $\varrho$ denotes a scaling factor to ensure that the total number of typical grids selected in the $t$-th subregion does not exceed that of all grids in it. It is noted from (\ref{a01}) that more typical grids should be selected if the size of the $t$-th subregion is large and/or its correlated range is small. It should also be mentioned that by varying the value of $\varrho$, there exists a fundamental trade-off between reducing the
power measurement overhead (for IRS channel estimation) and enhancing the coverage performance, as more/fewer typical grids will be selected by increasing/decreasing $\varrho$. It will be shown via simulation in Section \ref{sec007} that a small value of $\varrho$ usually suffices to achieve a good coverage performance.
Finally, we randomly select $\eta_t$ grids from the $t$-th subregion as the typical grids. Let ${\cal K}_2$ denote the set of all selected typical grids, with $\left|{\cal K}_2 \right| = K_2$. After this typical grid selection, the deterministic IRS channel estimation follows at the selected typical grids, as detailed in the next section.

\section{Single-layer NN-Based IRS Channel Estimation}\label{sec004}
In this section, we propose a NN-based channel estimation method based on the users' power measurements to estimate the deterministic IRS cascaded channel at each typical grid selected. For brevity, we present the proposed channel estimation for one specific typical grid (e.g., grid $k, k \in {\cal K}_2$), while it can be similarly applied to other typical grids.

\vspace{-9pt}

\subsection{NN-based Channel Estimation}\label{sec004B}


\begin{figure}[t]
\center
\includegraphics[scale=0.296]{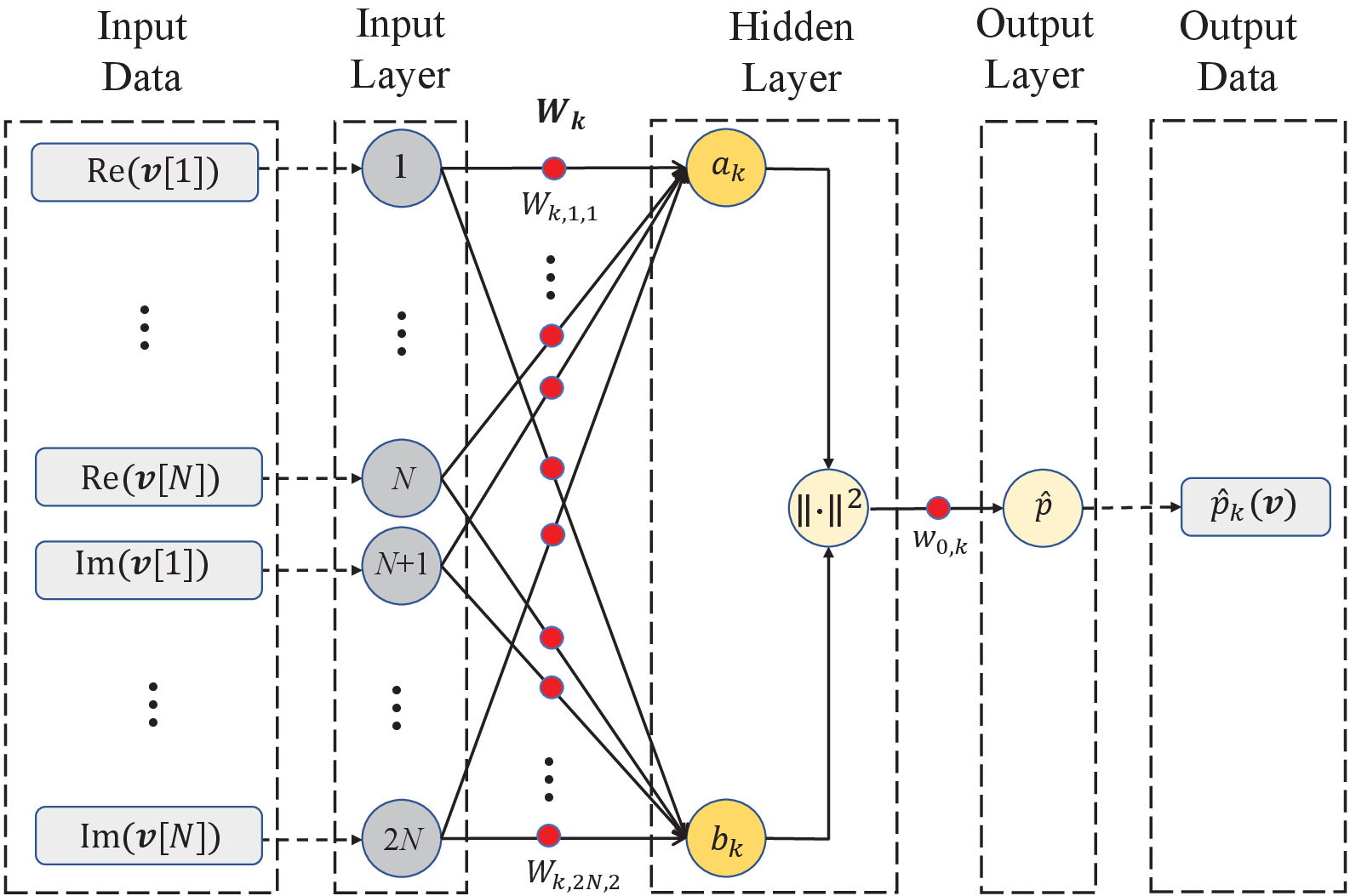}
\caption{Single-layer NN architecture for IRS channel estimation at grid $k$. }
\label{fig002}
\vspace{-10pt}
\end{figure}

Specifically, the mobile terminal in grid $k$ performs similar power measurement procedures as in Section \ref{sec004A}, but under a larger number of random IRS reflection sets (denoted as $M_2$, $M_2 \gg M_1$) in general. Under any given IRS reflection set $\boldsymbol{v}$ and with a sufficiently large $Q$, the average received signal power at gird $k$ will approach the expectation in (\ref{eqs0307}), i.e.,
\begin{equation}\label{eqs03007}
{\overline{p}}_k(\boldsymbol{v})={\left|{\boldsymbol{v}^H{\boldsymbol{\overline{h}}_k}} \right|}^2 + C_k.
\end{equation}
The NN-based method aims to estimate $\boldsymbol{\overline{h}}_k, k \in {{\cal{K}}_2}$, based on $\overline{p}_k(\boldsymbol{v}_m), m \in {\cal{M}}_2\triangleq \left\{1,2, \cdots ,M_2\right\}$. To this end, note that (\ref{eqs03007}) can be modeled as a single-layer NN, explained as follows. In particular, this NN takes the reflection pattern $\boldsymbol{v}$ and the cascaded channel ${\boldsymbol{\overline{h}}_k}$ as its input and weights, respectively. The nonlinear activation function at the output layer is the squared amplitude of ${\boldsymbol{v}^H{\boldsymbol{\overline{h}}_k}}$, followed by an additive weight to learn the unknown bias $C_{k}$, as given in (\ref{eqs03007}). However, as both $\boldsymbol{v}$ and ${\boldsymbol{\overline{h}}_k}$ are complex numbers in general, such a single-layer NN requires the implementation in the complex domain. To avoid this issue, we express (\ref{eqs03007}) equivalently in the real domain as
\begin{equation}\label{eqs03008}
{\overline{p}}_k(\boldsymbol{v})={\left\vert {\boldsymbol{v}^H{\boldsymbol{\overline{h}}_k}} \right\vert}^2 + C_k= \left\|{\boldsymbol{x}^T \boldsymbol{R}_k}\right\|^2 + C_k,
\end{equation}
where $\boldsymbol{x}$ consists of the real and imaginary parts of $\boldsymbol{v}$, i.e., $\boldsymbol{x}^T = \left[{\begin{array}{*{20}{c}}
{{\mathop{\rm Re}\nolimits} \left( {{\boldsymbol{v}^T}} \right)},&{{\mathop{\rm Im}\nolimits} \left( {{\boldsymbol{v}^T}} \right)}
\end{array}}\right]$, and $\boldsymbol{R}_k$ denotes the real-valued cascaded channel, i.e.,
\begin{equation}\label{eqs0300801}
\boldsymbol{R}_k=\left[ {\begin{array}{*{20}{c}}
{ {{\mathop{\rm Re}\nolimits} \left( \boldsymbol{\overline{h}}_k \right)} }&{{{{\mathop{\rm Im}\nolimits} \left( \boldsymbol{\overline{h}}_k \right)}}}\\
{ {{{\mathop{\rm Im}\nolimits} \left( \boldsymbol{\overline{h}}_k \right)}}}&{{-{{\mathop{\rm Re}\nolimits} \left( \boldsymbol{\overline{h}}_k \right)}}}
\end{array}} \right]\in \mathbb{R}^{{2N} \times 2}.
\end{equation}

Based on (\ref{eqs03008}), we can construct an equivalent single-layer NN to (\ref{eqs03007}) in the real-number domain. Specifically, as shown in Fig. \ref{fig002}, the input of this single-layer NN is $\boldsymbol{x}$. Let $W_{k,i,j}$ denote the weight of the edge from the $i$-th input to the $j$-th neuron in the hidden layer, with $i=1,2, \cdots ,2N$ and $j=1,2$. The two neurons at the hidden layer of this single-layer NN are given by
\begin{equation}\label{eqs03009}
  \left[ {\begin{array}{*{20}{c}}
{a_k}\\
{b_k}
\end{array}} \right]^T = \boldsymbol{x}^T \boldsymbol{W}_k,
\end{equation}
where $\boldsymbol{W}_k\in \mathbb{R}^{2N \times 2}$ denotes the multiplicative weight matrix of this NN, with $W_{k,i,j}$ being its entry in the $i$-th row and the $j$-th column.

Finally, the activation function is defined as the squared norm of (\ref{eqs03009}), and the output of this NN is given by
\begin{equation}\label{eqs03011}
  {{\hat p}_k}(\boldsymbol{v}) = a_k^2 + b_k^2 + {{w}}_{0, k} = \left\|{\boldsymbol{x}^T \boldsymbol{W}_k}\right\|^2 + {{w}}_{0, k},
\end{equation}
where ${{w}}_{0, k}$ denotes the additive weight for estimating the bias $C_k$.
By comparing (\ref{eqs03011}) with (\ref{eqs03008}), it is noted that this real-valued NN can imitate the received signal power at grid $k$. In particular, if ${\boldsymbol{R}}_k={\boldsymbol{W}}_k$ and $C_k={w}_{0, k}$, then we have $\hat p_k(\boldsymbol{v})=\overline{p}_k(\boldsymbol{v})$. Motivated by this, we propose to recover ${\boldsymbol{R}}_k$ (and ${\boldsymbol{\overline{h}}}_k$) by estimating the weight matrix ${\boldsymbol{W}}_k$ via training this single-layer NN. To this end, we consider that ${\boldsymbol{W}}_k$ takes a similar form to ${\boldsymbol{R}}_k$ in (\ref{eqs0300801}), i.e.,
\begin{equation}\label{eqs03010}
  {\boldsymbol{W}}_k=\left[ {\begin{array}{*{20}{c}}
{ {{\boldsymbol{w}_{1,k}}} }&{{{\boldsymbol{w}}_{2,k}}}\\
{ {{\boldsymbol{w}_{2,k}}}}&{{-{\boldsymbol{w}_{1,k}}}}
\end{array}} \right],
\end{equation}
where ${\boldsymbol{w}}_{1,k} \in \mathbb{R}^{N}$ and ${\boldsymbol{w}}_{2,k}\in \mathbb{R}^{N}$ correspond to ${{{\mathop{\rm Re}\nolimits} \left( \boldsymbol{\overline{h}}_k \right)} }$ and ${{{\mathop{\rm Im}\nolimits} \left( \boldsymbol{\overline{h}}_k \right)} }$ in (\ref{eqs0300801}), respectively.

With (\ref{eqs03010}), we present the following lemma.
\begin{lemma}\label{lemm001}
\textit{If $\hat p_k(\boldsymbol{v})=\overline{p}_k(\boldsymbol{v})$ holds for any $\boldsymbol{x} \in \mathbb{R}^{2N}$, we have ${\boldsymbol{\overline{h}}}_k={\boldsymbol{w}}_k e^{\jmath\xi_k}$ and $C_k={{w}}_{0, k}$, where ${\boldsymbol{w}}_k={{\boldsymbol{w}}_{1,k}} + \jmath {\boldsymbol{w}}_{2,k}$, and $\xi_k \in [0,2\pi)$ denotes an arbitrary phase.}
\end{lemma}

\begin{IEEEproof}
If $\hat p_k(\boldsymbol{v})=\overline{p}_k(\boldsymbol{v})$ holds for any $\boldsymbol{x} \in \mathbb{R}^{2N}$, we have
\begin{equation}\label{eqs03018}
  \|\boldsymbol{x}^T {\boldsymbol{W}}_k\|^2 + {{w}}_{0, k} = \|{\boldsymbol{x}^T \boldsymbol{R}_k}\|^2 + C_k.
\end{equation}
By substituting (\ref{eqs03010}) into the left-hand side of (\ref{eqs03018}), we have
\begin{equation}\label{eqs0301801}
  \|\boldsymbol{x}^T {\boldsymbol{W}}_k\|^2 = {\left| {\boldsymbol{v}^H{\boldsymbol{w}}_k} \right|}^2={\boldsymbol{v}^H{\boldsymbol{w}}_k}{{\boldsymbol{w}}_k^H\boldsymbol{v}} .
\end{equation}
Next, by substituting (\ref{eqs03008}) and (\ref{eqs0301801}) into (\ref{eqs03018}), we have
\begin{equation}\label{eqs03019}
{\boldsymbol{v}^H{\boldsymbol{w}}_k}{{\boldsymbol{w}}_k^H\boldsymbol{v}} + {{w}}_{0, k} ={{\boldsymbol{v}}^H{{{{\boldsymbol{\overline{h}}}}}_k{{{\boldsymbol{\overline{h}}}}}_k^H}{\boldsymbol{v}}}+ C_k, \ \forall \boldsymbol{v} \in {\mathbb{C}}^{N},
\end{equation}
which implies that
\begin{equation}\label{eqs03021}
{\boldsymbol{v}}^H\left({{{\boldsymbol{\overline{h}}}}}_k{{{\boldsymbol{\overline{h}}}}}_k^H - {\boldsymbol{w}}_k{\boldsymbol{w}}_k^H\right){\boldsymbol{v}} + \left({{w}}_{0, k} - C_k\right) = 0, \ \forall {\boldsymbol{v}} \in {\mathbb{C}}^{N}.
\end{equation}
For (\ref{eqs03021}) to hold for any ${\boldsymbol{v}} \in {\mathbb{C}}^{N}$, it should be satisfied that ${{{\boldsymbol{\overline{h}}}}}_k{{{\boldsymbol{\overline{h}}}}}_k^H = {\boldsymbol{w}}_k{\boldsymbol{w}}_k^H$ and $C_k={{w}}_{0, k}$. As such, we have ${\boldsymbol{\overline{h}}}_k={\boldsymbol{w}}_k e^{\jmath\xi_k}$. The proof is thus completed.
\end{IEEEproof}

It follows from Lemma \ref{lemm001} that we can estimate ${\boldsymbol{\overline{h}}}_k$ by training the single-layer NN in Fig. \ref{fig002} to estimate ${\boldsymbol{W}}_k$ and ${w}_{0,k}$ first. Although we cannot derive the exact ${\boldsymbol{\overline{h}}}_k$ due to the presence of the unknown phase $\xi_k$, the objective function of (P2) only depends on the channel covariance matrix ${\boldsymbol{{G}}}_k$, and we have ${\boldsymbol{{G}}}_k={{{\boldsymbol{\overline{h}}}}}_k{{{\boldsymbol{\overline{h}}}}}_k^H = {\boldsymbol{w}}_k{\boldsymbol{w}}_k^H, k \in {{\cal{K}}_2}$. As such, the unknown common phase does not affect the objective function of (P2). It should also be mentioned that Lemma \ref{lemm001} requires that (\ref{eqs03018}) holds for any ${\boldsymbol{x}} \in {\mathbb{R}}^{2N}$ or ${\boldsymbol{v}} \in {\mathbb{C}}^{N}$. However, due to (\ref{Zb}), the discrete IRS passive reflection set can only take a finite number of values in a subspace of ${\mathbb{C}}^{N}$, and ${\boldsymbol{\overline{h}}}_k={\boldsymbol{w}}_k e^{\jmath\xi_k}$ may not always hold in such a subspace. Nonetheless, the proposed design is still effective, as will be explained in Remark \ref{remark1} later.
\vspace{-16.3pt}

\subsection{NN Training }

To estimate ${{{\boldsymbol{w}}}}_k$ or ${{{\boldsymbol{W}}}}_k$ (and ${{w}}_{0, k}$), we can train the NN in Fig. \ref{fig002} by using the stochastic gradient descent method to minimize the MSE between its output and the training data.
In particular, we can make full use of the power measurements, i.e., $\boldsymbol{p}_k, k \in {\cal{K}}_2$, as the training data. Specifically, we divide them into two data sets, namely, the training set and validation set. The training set consists of $M_0$ $(M_0<M_2)$ entries of $\boldsymbol{p}_k$, while the remaining $M_2-M_0$ entries of $\boldsymbol{p}_k$ are used as the validation set to evaluate the model fitting accuracy. As such, the MSE for the training data is set as the following loss function,
\begin{equation}\label{eqs0310}
 {\cal L}\left({\boldsymbol{{W}}_k,{{w}}_{0, k}}\right) = \frac{1}{M_0}\sum_{m=1}^{M_0}\left(\left\|{\boldsymbol{x}_m^T \boldsymbol{{W}}_k}\right\|^2 + {{w}}_{0, k} - \overline{p}_k(\boldsymbol{v}_m)\right)^2.
\end{equation}
According to Lemma \ref{lemm001}, the acquisition of $\boldsymbol{\overline{h}}_k$ requires to minimize the above MSE via updating the NN weights ${\boldsymbol{W}}_k$ and ${{w}}_{0, k}$. To this end, we first estimate $w_{0, k}$ by taking the partial derivative of (\ref{eqs0310}) to it, i.e.,
\begin{equation}\label{eqs4013}
  \frac{\partial{{\cal L}\left({\boldsymbol{{W}}_k,{{w}}_{0, k}}\right)}}{\partial{{{w}}_{0, k}}} = \frac{2}{M_0}\sum_{m=1}^{M_0}\left( {{w}}_{0, k} + \left\|{\boldsymbol{x}_m^T \boldsymbol{{W}}_k}\right\|^2 - \overline{p}_k(\boldsymbol{v}_m)\right).
\end{equation}
By setting (\ref{eqs4013}) to zero, we can obtain the estimate of ${{w}}_{0, k}$ as
\begin{equation}\label{eqs4014}
  {w}^{\star}_{0, k} = \frac{1}{M_0}\sum_{m=1}^{M_0}\left( \overline{p}_k(\boldsymbol{v}_m) - \left\|{\boldsymbol{x}_m^T {\boldsymbol{W}}_k}\right\|^2\right),
\end{equation}
which is a function of ${\boldsymbol{W}}_k$. By substituting (\ref{eqs4014}) into (\ref{eqs0310}), the loss function can be simplified as a function of $\boldsymbol{{W}}_k$, i.e.,
\begin{equation}\label{eqs4015}
  {\cal L}\left({\boldsymbol{{W}}_k}\right) = \frac{1}{M_0}\sum_{m=1}^{M_0}\left(\left\|{\boldsymbol{x}_m^T \boldsymbol{{W}}_k}\right\|^2 + {w}^\star_{0,k} - \overline{p}_k(\boldsymbol{v}_m)\right)^2.
\end{equation}

Given this loss function, we can use the backward propagation \cite{rumelhart1986learning} to iteratively update the NN weights. Specifically, with (\ref{eqs03010}), the weight matrix ${\boldsymbol{W}_k}$ can be expressed by a vector $\boldsymbol{\gamma}_k =\left[{{\boldsymbol{w}}_{1,k}}^T, {\boldsymbol{w}}_{2,k}^T\right]^T \in \mathbb{R}^{2N}$. Let $\boldsymbol{\gamma}_{k,z}$ denote the updated value of $\boldsymbol{\gamma}_k$ after the $z$-th iteration. The iteration proceeds as
\begin{equation}\label{eqs0312}
  {\boldsymbol{\gamma}_{ k , z + 1 }} = {\boldsymbol{\gamma}_{k,z}} - \kappa F\left({\boldsymbol{\gamma}_{k,z}}\right),
\end{equation}
where $\kappa>0$ denotes the learning rate, and $F\left({\boldsymbol{\gamma}_{k}}\right)=\nabla_{{\boldsymbol{\gamma}}_k}{\cal L}\left({\boldsymbol{{W}}_k}\right)$ denotes the gradient of the loss function ${\cal L}\left({\boldsymbol{{W}}_k}\right)$ with respect to $\boldsymbol{\gamma}_k$.
Here, $F\left({\boldsymbol{\gamma}_{k}}\right)$ can be computed using the chain rule,
\begin{equation}\label{eqs000001}
  F\left(\boldsymbol{\gamma}_k\right) = \frac{\partial{{\cal L}\left({\boldsymbol{{W}}_k}\right)}}{\partial{{{\hat{p}}_k}}}\left[\frac{\partial{{{\hat{p}}_k}}}{\partial{{{{a}}_k}}}, \ \frac{\partial{{{\hat{p}}_k}}}{\partial{{{{b}}_k}}}\right]
  \left[\nabla_{{{\boldsymbol{\gamma}}}_k}{{{a}}_k}, \ \nabla_{{{\boldsymbol{\gamma}}}_k}{{{b}}_k}\right]^T,
\end{equation}
where $\frac{\partial{{\cal L}\left({\boldsymbol{{W}}_k}\right)}}{\partial{{{\hat{p}}_k}}}$ can be computed based on (\ref{eqs4015}), while the other partial derivatives in (\ref{eqs000001}) can be calculated based on (\ref{eqs03011}) as
\begin{align*}
\frac{\partial{{{\hat{p}}_k}}}{\partial{{{{a}}_k}}} & = 2{{{a}}_k}, \ \text{and} \ \
\frac{\partial{{{\hat{p}}_k}}}{\partial{{{{b}}_k}}} = 2{{b}}_k.
\end{align*}
Note that the two gradients in (\ref{eqs000001}) can be computed based on (\ref{eqs03009}) as
\begin{align}
\nabla_{{{\boldsymbol{\gamma}}}_k}{{{a}}_k} & = \left[ {1,
{\cos{{\theta _1}} },\!\cdots\!, {\cos {{\theta _{N}}}},
{0, -\sin{{\theta _1}}},\!\cdots\!,{-\sin{{\theta _{{N}}}}}} \right]^T,\nonumber\\
\nabla_{{{\boldsymbol{\gamma}}}_k}{{{b}}_k} & = \left[
{0, \sin{{\theta_1}}},\! \cdots\!,{\sin{{\theta_{N}}}},
{1, \cos{{\theta_1}}},\! \cdots\!,{\cos{{\theta_{N}}}} \right]^T\nonumber.
\end{align}
The NN training process terminates after $Z$ rounds of iterations, and the weight matrix of the NN is determined as
\begin{equation}\label{eqs0311}
  \boldsymbol{W}^{\star}_k = \arg \min_{1 \le z \le Z} \left(\sum_{m=M_0+1}^{M_2} {(\hat {p}_{k,z}}(\boldsymbol{v}_m)-{{\overline{p}}_k}(\boldsymbol{v}_m))^2\right),
\end{equation}
based on the validation set, where ${\hat{p}}_{k,z}(\boldsymbol{v}_m)=\|\boldsymbol{x}_m^T {\boldsymbol{W}}_{k,z}\|^2$ denotes the output of the NN after the $z$-th iteration, and $\boldsymbol{W}_{k,z}$ denotes the updated version of $\boldsymbol{W}_k$ after the $z$-th iteration with $z = 1,2, \cdots ,Z$. Based on the above, the complex-valued cascaded channel can be estimated as ${\boldsymbol{w}}^{\star}_k={{\boldsymbol{w}}^{\star}_{1,k}} + \jmath {\boldsymbol{w}}^{\star}_{2,k}$, and the estimated channel covariance is denoted as $ \boldsymbol{{\hat G}}_k = \boldsymbol{w}_k^\star \boldsymbol{w}_k^{\star H}$.

\begin{remark}\label{remark1}
In the special case with one-bit IRS phase shifts, i.e., $\alpha =1$, the cascaded channel ${\boldsymbol{\overline{h}}}_k$ may not be estimated as ${\boldsymbol{w}}_k e^{\jmath\xi_k}$. This is because in this case, we have ${\boldsymbol{v}^*}={\boldsymbol{v}}$, which results in
\begin{equation}\label{eqs03023}
  {\boldsymbol{v}^H}{{{\boldsymbol{\overline{h}}}}}_k{{{\boldsymbol{\overline{h}}}}}_k^H\boldsymbol{v}={\boldsymbol{v}^H}{\boldsymbol{\overline{h}}}_k^{*}{\boldsymbol{\overline{h}}}_k^{T}\boldsymbol{v}.
\end{equation}
Based on (\ref{eqs03021}), we may estimate ${\boldsymbol{\overline{h}}}_k^*$ as ${\boldsymbol{w}}_k e^{\jmath\xi_k}$, while the actual channel ${\boldsymbol{\overline{h}}}_k$ is ${\boldsymbol{w}}_k^* e^{-\jmath\xi_k}$. However, this does not affect the efficacy of the proposed design, since both estimations lead to the same received signal power due to (\ref{eqs03023}).
\end{remark}

In the proposed IRS channel estimation method, the computational complexity is mainly due to the NN training procedures. In particular, the training complexity depends on the size of the NN structure. In the NN for grid $k$, as shown in Fig. \ref{fig002}, the number of neurons is $2$, and the number of weights is $2N+1$, which incurs the computational complexity for each grid in the order of ${\cal O}\left(N\right)$\cite{haykinneural}.

Note that as compared to the conventional cascaded channel estimation method requiring additional pilot signals, the proposed method only requires power measurements taken at each grid, which are easily accessible in the current wireless communication systems. Based on the channel estimates,  the IRS passive reflection can be optimized by maximizing the average received signal power over the selected grids in ${\cal K}_2$ via replacing ${\cal K}$ and $K$ in (P2) with ${\cal K}_2$ and $K_2$, respectively. We will evaluate the performance of different algorithms for solving this problem in Section \ref{sec007} via simulation.

\section{Numerical Results}\label{sec007}

In this section, numerical results are presented to examine the performance of our proposed design framework with power measurement location selection, IRS channel estimation and IRS passive reflection for coverage enhancement under the practical ray-tracing based channel model.

\subsection{Simulation Setup}


We focus on a real-world simulation setup, i.e., a square subregion in London, where the locations of the BS and the IRS are marked in red. We assume that the central (reference) point of the BS is deployed at a latitude of $51.5031$ degrees north and a longitude of $0.0165$ degrees west at the height of $20$ meters (m), while the IRS is deployed at a latitude of $51.50313$ degrees north and a longitude of $0.0178$ degrees west at the height of $1$ m. The region $\cal D$ of interest is a square with a side length of $30$ m. The reference grid of this region (the bottom-left grid) is located at the point with a latitude of $51.50279$ degrees north and a longitude of $0.0180$ degrees west at the height of $1$ m above the ground. This region is uniformly divided into $K = 10 \times 10$ grids, each with a size of 3 m $\times$ 3 m. Moreover, the IRS is equipped with a UPA comprising $N=8 \times 8$ reflecting elements with half-wavelength spacing. The number of control bits for the IRS phase shifts is set to $\alpha=2$, if not specified otherwise.


\begin{figure}[t]
\begin{center}
\centering
\includegraphics[scale=0.53601369]{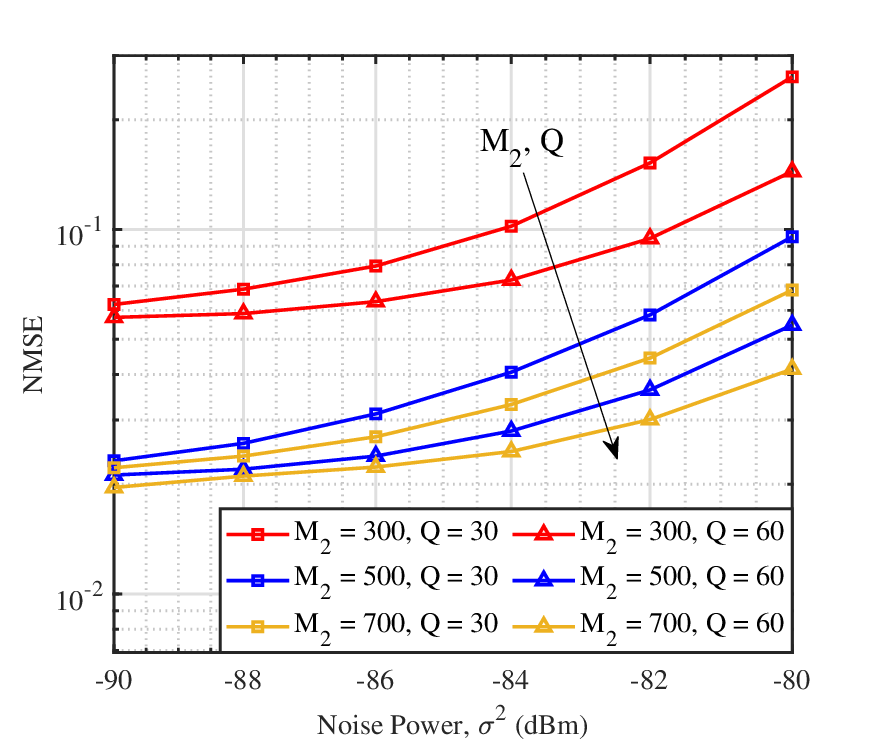} 
\caption{NMSE of IRS channel estimation versus the receiver noise power.}
\label{fig07004}
\end{center}
\vspace{-16pt}
\end{figure}

\subsection{IRS Channel Estimation }

To evaluate the performance of the proposed single-layer NN-based IRS channel estimation method, we first consider that all grids in $\cal D$ are selected for channel estimation, i.e., $K_2=K$, so as to eliminate the effects of power-measurement grid selection. In particular, the stochastic gradient descent method is utilized to perform the NN training, employing a learning rate of $2 \times 10^{-2}$ that decays $1 \times 10^{-4}$ per epoch. The batch size is set to $2$, and there are $200$ epochs in total. To evaluate the channel estimation accuracy, we calculate the average normalized mean square error (NMSE) between the estimated channel covariance matrices (i.e., $\boldsymbol{\hat{{G}}}_{k}$) and the actual ones (i.e., $\boldsymbol{{G}}_{k}$) over all $K$ grids in the region $\cal D$, which is given by
\begin{equation}
   \text{NMSE} = \frac{1}{K}\sum_{k \in \cal{K}}{\frac{\left\|\boldsymbol{\hat{{G}}}_{k} - \boldsymbol{{{G}}}_{k}\right\|^2_F}{\left\|\boldsymbol{{{G}}}_{k}\right\|^2_F}}.
\end{equation}

In Fig. \ref{fig07004}, we show the NMSE by the proposed IRS channel estimation method versus the noise power under different numbers of random IRS reflection sets, i.e., $M_2=300, 500, 700$, and different numbers of received symbols per reflection set, i.e., $Q=30,60$. It is observed that the NMSE decreases with $M_2$ thanks to the enlarged training data set even with a high noise power, which shows the robustness of our proposed single-layer NN-based IRS channel estimation against receiver noise.
Furthermore, it is observed that the estimation accuracy improves with increasing $Q$. This is expected as with a large $Q$, the random channel components and noise can be more effectively averaged out, such that the power measurement can better approximate the received signal power due to deterministic IRS channels in (\ref{eqs03007}) only.

\subsection{Coverage Performance}

Next, assuming that all grids are selected for IRS channel estimation again, we evaluate the average received SNR (in dB) by different schemes, i.e.,
\begin{equation}
  \text{SNR} = 10\log_{10}{\frac{\sum_{{{k \in \mathcal{K}}}} \ \left(\boldsymbol{v}^{\star H}\boldsymbol{{G}}_{k}\boldsymbol{v}^{\star} + C_k - \sigma^2\right)}{K\sigma^2}},
\end{equation}
where $\boldsymbol{v}^{\star}$ denotes the optimized IRS passive reflection for coverage enhancement by different schemes.

Specifically, to obtain $\boldsymbol{v}^{\star}$ in our proposed framework, we solve (P2) with the estimated CSI by replacing $\boldsymbol{{G}}_k$ therein with $\boldsymbol{\hat{G}}_k$. Although such a problem is non-convex, it can be relaxed into
a semi-definite programming (SDP) problem by assuming continuous IRS phase shifts, for which the Gaussian randomization and solution quantization can be applied to reconstruct a feasible solution satisfying constraint (\ref{p2c}). Next, we successively refine the phase shifts of the reconstructed solution until convergence. The details of solving (P2) can be found in \cite{sun} and omitted in this paper for brevity.

Moreover, we adopt the CSM \cite{CSM,ACSM} and random-max sampling (RMS)\cite{CSM,tao2020intelligent} methods as benchmark schemes, both of which also design the IRS passive reflection based on power measurements, but without estimating the IRS cascaded channels as our proposed method. Specifically, the RMS method sets the IRS reflection as the one that yields the maximum average received signal power over all selected grids among $M_2$ random IRS training reflection sets, i.e.,
\begin{equation}\label{s3001}
  {\boldsymbol{v}}^{\text{RMS}} = \boldsymbol{v}_{m^{\star}}, \ \ \ \text{with} \ \ \ {m^{\star}}=\arg{{\max_{m \in {\mathcal{M}_2}} \sum_{k \in \mathcal{K}}{\frac{1}{K}{\bar{p}}_{k}(\boldsymbol{v_m})}}}.
\end{equation}

The CSM method in \cite{CSM} calculates the sample mean of power measurements over all selected power-measurement locations conditioned on $\theta_i=\psi, \psi \in \Phi_{\alpha}$, i.e.,
\begin{equation}\label{s3002}
  {{\mathbb{E}}}[p|\theta_i=\psi] = \frac{1}{\left| \mathcal{S}_{i}(\psi) \right|} \sum_{\boldsymbol{v} \in \mathcal{S}_{i}(\psi)}{\sum_{k \in \mathcal{K}}{\frac{1}{K }{\bar{p}}_{k}(\boldsymbol{v})}},
\end{equation}
where $\mathcal{S}_{i}(\psi)$ denotes a subset of the $M_2$ random reflection sets with $\theta_i=\psi$, $i=1,2, \cdots ,N$. Finally, the phase shift of the $i$-th reflecting element is set as
\begin{equation}\label{s3003}
  \theta_i^{\text{CSM}} = \arg{\max_{\psi \in \Phi_\alpha}{{{\mathbb{E}}}[p|\theta_i=\psi]}}, \ i=1, \cdots ,N.
\end{equation}
However, the authors in \cite{CSM} identified that the above CSM may not perform well in the case without a direct BS-user link. As such, they proposed an improved adaptive CSM (ACSM) algorithm in \cite{ACSM} by treating the cascaded channel via a subset of IRS reflecting elements as a virtual direct BS-user link and then applying the above CSM. Hence, we adopt this ACSM algorithm as a benchmark in this paper, and interested readers can refer to \cite{ACSM} for its details.

\begin{figure}[t]
  \centering
  \subfigure[$\alpha=1$]{\includegraphics[width=0.41361035\textwidth]{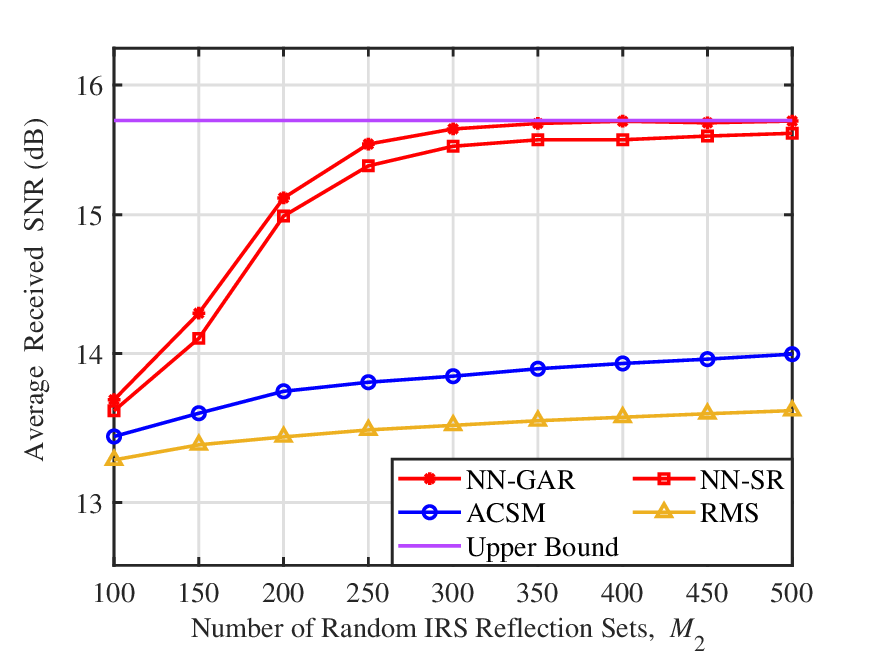}}  
  \subfigure[$\alpha=2$]{\includegraphics[width=0.4163135\textwidth]{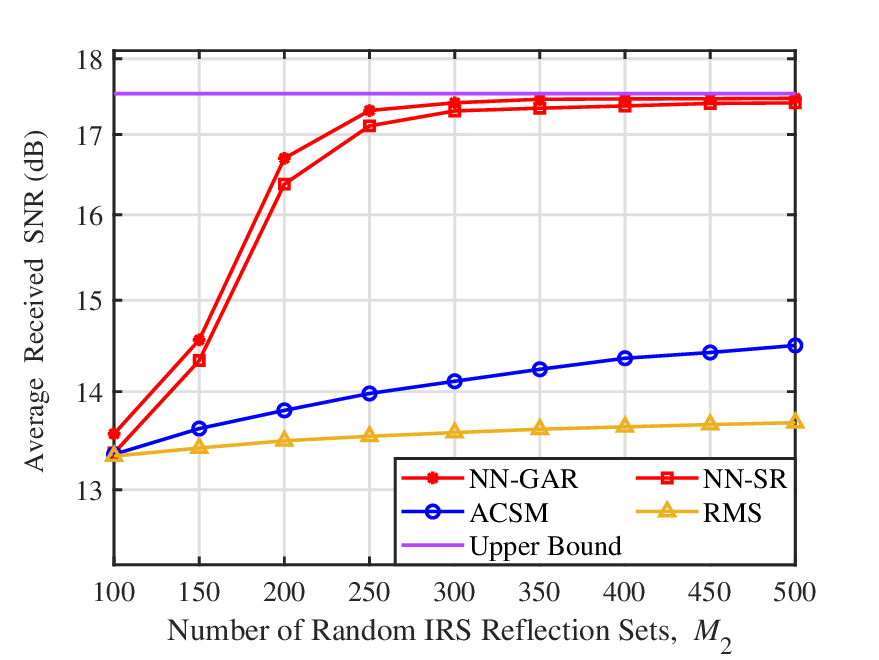}}  
  \caption{Average received SNR versus the number of IRS reflection sets by different schemes.}
\label{fig07007}
\end{figure}

Finally, the IRS passive reflection design based on perfect CSI for all grids (i.e., by solving (P2) under perfect knowledge of ${{\boldsymbol{G}}}_k, k \in {\cal{K}}$) is included as a performance upper bound for our proposed algorithms (labeled as ``Upper Bound'').

Figs. \ref{fig07007}(a) and \ref{fig07007}(b) plot the average received SNRs by different schemes versus the number of IRS reflection sets for channel estimation, $M_2$, with $Q=60$ and the number of control bits for IRS phase shifts set to $\alpha=1$ and $\alpha=2$, respectively. The noise power is $\sigma^2=-90$ dBm. For the proposed IRS passive reflection design, in addition to the aforementioned Gaussian randomization method (labeled as ``NN-GAR''), we also show the performance by directly applying the successive refinement method, where the IRS passive reflection is initialized based on the RMS given in (\ref{s3001}) (labeled as ``NN-SR''). Note that as compared to NN-GAR with polynomial complexity over $N$ [1], NN-SR only incurs linear complexity over $N$. It is observed that both NN-GAR and NN-SR significantly outperform the benchmark schemes by fully exploiting the power measurements for channel estimation. In particular, with increasing $M_2$, the performance of our proposed scheme becomes closer to the upper bound under perfect CSI at each grid. Moreover, the average SNR performance improves by increasing $\alpha$ from 1 to 2, as expected, thanks to the higher phase-shift resolution for both channel estimation and reflection design. The above observations demonstrate that the IRS passive reflection can be efficiently optimized for coverage enhancement with linear channel estimation complexity over $N$ based on the power measurements by utilizing the NN-SR scheme.

\subsection{Effect of Power-Measurement Locations}

\begin{figure*}[t]
\begin{center}
\centering
\includegraphics[scale=0.569019636]{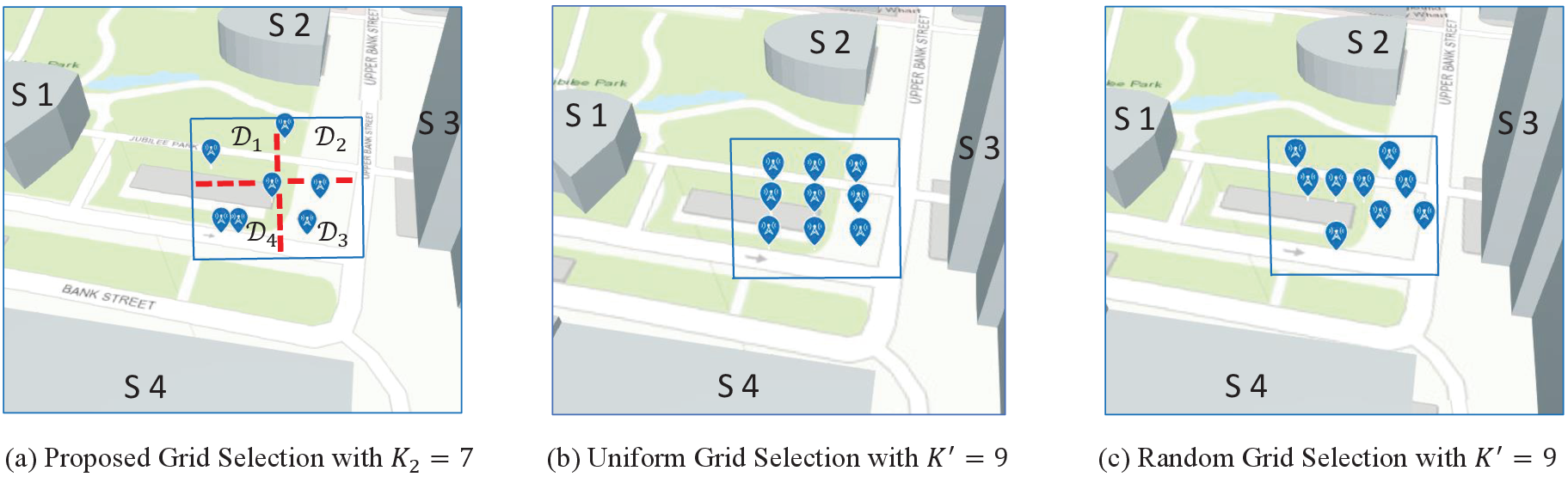} 
\caption{Power-measurement grids selected by different algorithms.}
\label{fig07002}\end{center}
\vspace{-16.9pt}
\end{figure*}

Next, we show the average received SNRs achieved by the proposed framework under different grid selection algorithms. In our proposed typical grid selection method, the user terminal randomly selects $K_1=20$ grids to conduct initial power measurements, and we set the number of IRS reflection sets to $M_1=10$, with $Q=60$ received symbols per reflection set. Moreover, we use the spherical variogram as the empirical model for variogram estimation since it is a generalized model that incorporates other common variogram models, e.g., the exponential variogram model and the Gaussian variogram model\cite{chi1999}. As shown in Fig. \ref{fig07002}(a), the considered region ${\cal D}$ is split into $T=4$ subregions (labeled as ``${\cal D}_1$'', ``${\cal D}_2$'', ``${\cal D}_3$'' and ``${\cal D}_4$'' in a clockwise manner). It is observed that several dominant scatterers are located in the vicinity of each subregion, i.e., scatterers ``S1'' and ``S2'' for subregion ``${\cal D}_1$'', scatterers ``S2'' and ``S3'' for subregion ``${\cal D}_2$'', scatter ``S3'' for subregion ``${\cal D}_3$'', and scatterers ``S1'' and ``S4'' for subregion ``${\cal D}_4$'', resulting in heterogeneous propagation environments for them. The correlated ranges of the four subregions are obtained as $c^{\star}_1 = 11.79$ m, $c^{\star}_2 = 12.73$ m, $c^{\star}_3 = 8.93$ m and $c^{\star}_4 = 6.11$ m, based on which the numbers of typical grids in the four subregions can be obtained as $\eta_1=1$, $\eta_2=1$, $\eta_3=2$ and $\eta_4=3$, respectively, by setting $\varrho=0.4$. Hence, there are ${K}_2=7$ grids selected in total for the subsequent power measurements and IRS channel estimation, which is considerably smaller than the total number of grids in $\cal D$, i.e., $K=100$.

To evaluate the efficacy of our proposed grid selection algorithm, we consider the uniform selection and random selection as benchmark schemes. Let $K'$ denote the number of power-measurement grids selected by these two benchmark schemes. For the uniform selection, we uniformly divide the region $\cal D$ into $K'$ subregions, and the grids around their centers are selected, as shown in Fig. \ref{fig07002}(b) with $K'=9$. While for the random selection scheme, we randomly select $K'$ grids in $\cal D$, as shown in Fig. \ref{fig07002}(c) with $K'=9$.

Fig. \ref{fig07003} shows the average received SNR after IRS passive reflection optimization by different grid selection schemes versus the number of selected power-measurement grids in the two benchmark algorithms, i.e., $K'$, with $\sigma^2 = - 90$ dBm. Note that we fix $K_2=7$ and $K'$ is set as a square number in Fig. \ref{fig07003} to realize the uniform division of the square region $\cal D$. In all considered location selection schemes, the IRS cascaded channels at the selected grids are estimated by our proposed IRS channel estimation algorithm for IRS passive reflection design with $M_2=600$ and $Q=60$. It is observed that the proposed grid selection scheme can achieve better average SNR performance than the two benchmark schemes even with much fewer grids selected (i.e., ${K}_2=7$ versus $K'=25$) as well as a close performance to the upper bound assuming full and perfect CSI. All of the above results in this section demonstrate the efficacy of our proposed framework in terms of both grid selection and IRS channel estimation for coverage performance enhancement. 

\begin{figure}[t]
\begin{center}
\centering
\includegraphics[scale=0.5169]{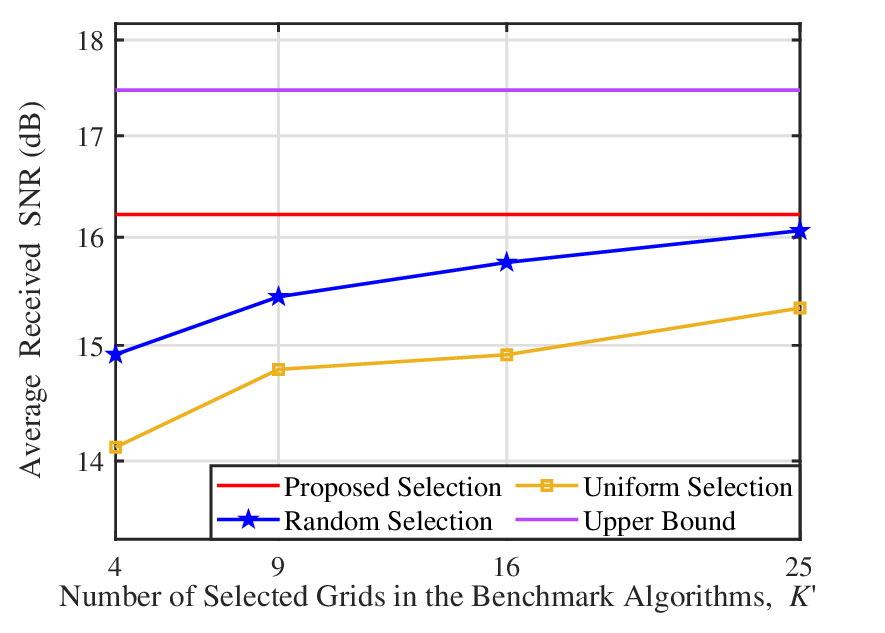} 
\caption{Average received SNR by different location selection algorithms versus number of selected power-measurement grids in the benchmark algorithms.}
\vspace{-16pt}\label{fig07003}\end{center}
\end{figure}

\section{Conclusion}\label{sec008}

This paper proposed an IRS-aided coverage performance optimization framework for a given region by exploiting the offline power measurements. To efficiently acquire the deterministic IRS cascaded channel information required for IRS reflection optimization, a grid selection method was first designed to select only a small number of typical grids for IRS channel estimation, by exploiting the inherent channel spatial correlation in the considered region. Moreover, to estimate the deterministic IRS cascaded channel at each selected grid, we proposed a single-layer NN-based channel estimation method based on the power measurements with randomly generated IRS reflections without the need for additional pilot transmission. Numerical results based on the practical ray-tracing channel model showed that our proposed framework can significantly outperform the existing power-measurement-based schemes and heuristic grid selection schemes, and approach the performance upper bound assuming perfect CSI. Our proposed IRS-assisted coverage enhancement framework can be extended to more general system setups, such as multi-antenna BS with direct user channels, wideband communications, and multi-IRS reflections, which are worthy of investigation in future work. How to select power-measurement locations for channel estimation in a given environment with minimum overhead is also worth further studying.

where ${a_{1,k}}, {a_{2,k}}, {a_{3,k}}$ and ${a_{4,k}}$ are given by (\ref{eqs03005}).

\bibliographystyle{IEEEtran}
\bibliography{reference}

\begin{thebibliography}{10}
\providecommand{\url}[1]{#1}
\csname url@samestyle\endcsname
\providecommand{\newblock}{\relax}
\providecommand{\bibinfo}[2]{#2}
\providecommand{\BIBentrySTDinterwordspacing}{\spaceskip=0pt\relax}
\providecommand{\BIBentryALTinterwordstretchfactor}{4}
\providecommand{\BIBentryALTinterwordspacing}{\spaceskip=\fontdimen2\font plus
\BIBentryALTinterwordstretchfactor\fontdimen3\font minus
  \fontdimen4\font\relax}
\providecommand{\BIBforeignlanguage}[2]{{%
\expandafter\ifx\csname l@#1\endcsname\relax
\typeout{** WARNING: IEEEtran.bst: No hyphenation pattern has been}%
\typeout{** loaded for the language `#1'. Using the pattern for}%
\typeout{** the default language instead.}%
\else
\language=\csname l@#1\endcsname
\fi
#2}}
\providecommand{\BIBdecl}{\relax}
\BIBdecl

\bibitem{sun}
H.~Sun, W.~Mei, L.~Zhu, and R.~Zhang, ``User power measurement based {IRS}
  channel estimation via single-layer neural network,'' \emph{arXiv:2309.08275,
  to appear in Proc. IEEE Global Commun. Conf. (GLOBECOM)}, Kuala Lumpur,
  Malaysia, Dec. 2023.

\bibitem{WQQTutorial}
Q.~Wu, S.~Zhang, B.~Zheng, C.~You, and R.~Zhang, ``Intelligent reflecting
  surface-aided wireless communications: A tutorial,'' \emph{IEEE Trans.
  Commun.}, vol.~69, no.~5, pp. 3313--3351, May 2021.

\bibitem{wu2023intelligent}
Q.~Wu, B.~Zheng, C.~You, L.~Zhu, K.~Shen, X.~Shao, W.~Mei, B.~Di, H.~Zhang,
  E.~Basar, L.~Song, M.~D. Renzo, Z.-Q. Luo, and R.~Zhang, ``Intelligent
  surfaces empowered wireless network: Recent advances and the road to {6G},''
  \emph{arXiv preprint arXiv:2312.16918}, 2023.

\bibitem{surveyapp}
S.~Gong, X.~Lu, D.~T. Hoang, D.~Niyato, L.~Shu, D.~I. Kim, and Y.-C. Liang,
  ``Toward smart wireless communications via intelligent reflecting surfaces: A
  contemporary survey,'' \emph{IEEE Commun. Surveys Tuts.}, vol.~22, no.~4, pp.
  2283--2314, 4th Quart. 2020.

\bibitem{mei}
W.~Mei, B.~Zheng, C.~You, and R.~Zhang, ``Intelligent reflecting surface-aided
  wireless networks: From single-reflection to multireflection design and
  optimization,'' \emph{Proc. IEEE}, vol. 110, no.~9, pp. 1380--1400, Sep.
  2022.

\bibitem{IRSMIMO}
S.~Zhang and R.~Zhang, ``Capacity characterization for intelligent reflecting
  surface aided {MIMO} communication,'' \emph{IEEE J. Sel. Areas Commun.},
  vol.~38, no.~8, pp. 1823--1838, Aug. 2020.

\bibitem{IRSMIMO2}
K.~Zhi, C.~Pan, H.~Ren, and K.~Wang, ``Statistical {CSI}-based design for
  reconfigurable intelligent surface-aided massive {MIMO} systems with direct
  links,'' \emph{IEEE Wireless Commun. Lett.}, vol.~10, no.~5, pp. 1128--1132,
  May 2021.

\bibitem{JSAC22}
X.~Shao, C.~You, W.~Ma, X.~Chen, and R.~Zhang, ``Target sensing with
  intelligent reflecting surface: Architecture and performance,'' \emph{IEEE J.
  Sel. Areas Commun.}, vol.~40, no.~7, pp. 2070--2084, Jul. 2022.

\bibitem{10149442}
X.~Pang, W.~Mei, N.~Zhao, and R.~Zhang, ``Cellular sensing via cooperative
  intelligent reflecting surfaces,'' \emph{IEEE Trans. Veh. Technol.}, vol.~72,
  no.~11, pp. 15\,086--15\,091, Nov. 2023.

\bibitem{UAV}
C.~You, Z.~Kang, Y.~Zeng, and R.~Zhang, ``Enabling smart reflection in
  integrated air-ground wireless network: {IRS} meets {UAV},'' \emph{IEEE
  Wireless Commun.}, vol.~28, no.~6, pp. 138--144, Dec. 2021.

\bibitem{9777746}
X.~Pang, W.~Mei, N.~Zhao, and R.~Zhang, ``Intelligent reflecting surface
  assisted interference mitigation for cellular-connected {UAV},'' \emph{IEEE
  Wireless Commun. Lett.}, vol.~11, no.~8, pp. 1708--1712, Aug. 2022.

\bibitem{pls}
X.~Yu, D.~Xu, Y.~Sun, D.~W.~K. Ng, and R.~Schober, ``Robust and secure wireless
  communications via intelligent reflecting surfaces,'' \emph{IEEE J. Sel.
  Areas Commun.}, vol.~38, no.~11, pp. 2637--2652, Nov. 2020.

\bibitem{IRSOFDM}
Y.~Yang, B.~Zheng, S.~Zhang, and R.~Zhang, ``Intelligent reflecting surface
  meets {OFDM}: Protocol design and rate maximization,'' \emph{IEEE Trans.
  Commun.}, vol.~68, no.~7, pp. 4522--4535, Jul. 2020.

\bibitem{IRSOTFS}
M.~Li, S.~Zhang, Y.~Ge, F.~Gao, and P.~Fan, ``Joint channel estimation and data
  detection for hybrid {RIS} aided millimeter wave {OTFS} systems,'' \emph{IEEE
  Trans. Commun.}, vol.~70, no.~10, pp. 6832--6848, Oct. 2022.

\bibitem{cells}
W.~Mei and R.~Zhang, ``Performance analysis and user association optimization
  for wireless network aided by multiple intelligent reflecting surfaces,''
  \emph{IEEE Trans. Commun.}, vol.~69, no.~9, pp. 6296--6312, Sep. 2021.

\bibitem{IRSOMA}
B.~Zheng, Q.~Wu, and R.~Zhang, ``Intelligent reflecting surface-assisted
  multiple access with user pairing: {NOMA} or {OMA}?'' \emph{IEEE Commun.
  Lett.}, vol.~24, no.~4, pp. 753--757, Apr. 2020.

\bibitem{IRSCESurvey1}
A.~L. Swindlehurst, G.~Zhou, R.~Liu, C.~Pan, and M.~Li, ``Channel estimation
  with reconfigurable intelligent surfaces {---} {A} general framework,''
  \emph{Proc. IEEE}, vol. 110, no.~9, pp. 1312--1338, Sep. 2022.

\bibitem{IRSCESurvey2}
B.~Zheng, C.~You, W.~Mei, and R.~Zhang, ``A survey on channel estimation and
  practical passive beamforming design for intelligent reflecting surface aided
  wireless communications,'' \emph{IEEE Commun. Surveys Tuts.}, vol.~24, no.~2,
  pp. 1035--1071, 2nd Quart. 2022.

\bibitem{DDNN}
S.~Liu, Z.~Gao, J.~Zhang, M.~D. Renzo, and M.-S. Alouini, ``Deep denoising
  neural network assisted compressive channel estimation for {mmWave}
  intelligent reflecting surfaces,'' \emph{IEEE Trans. Veh. Technol.}, vol.~69,
  no.~8, pp. 9223--9228, Aug. 2020.

\bibitem{YWJSAC}
T.~Jiang, H.~V. Cheng, and W.~Yu, ``Learning to reflect and to beamform for
  intelligent reflecting surface with implicit channel estimation,'' \emph{IEEE
  J. Sel. Areas Commun.}, vol.~39, no.~7, pp. 1931--1945, Jul. 2021.

\bibitem{You1}
C.~You, B.~Zheng, and R.~Zhang, ``Channel estimation and passive beamforming
  for intelligent reflecting surface: Discrete phase shift and progressive
  refinement,'' \emph{IEEE J. Sel. Areas Commun.}, vol.~38, no.~11, pp.
  2604--2620, Nov. 2020.

\bibitem{You2}
------, ``Fast beam training for {IRS}-assisted multiuser communications,''
  \emph{IEEE Wireless Commun. Lett.}, vol.~9, no.~11, pp. 1845--1849, Nov.
  2020.

\bibitem{CSM}
S.~Ren, K.~Shen, Y.~Zhang, X.~Li, X.~Chen, and Z.-Q. Luo, ``Configuring
  intelligent reflecting surface with performance guarantees: Blind
  beamforming,'' \emph{IEEE Trans. Wireless Commun.}, vol.~22, no.~5, pp.
  3355--3370, May 2023.

\bibitem{ACSM}
W.~Wang, W.~Lai, S.~Ren, L.~Xiang, X.~Li, S.~Niu, and K.~Shen, ``Adaptive
  beamforming for non-line-of-sight {IRS}-assisted communications without
  {CSI},'' in \emph{Proc. IEEE Ann. Int. Symp. Personal Indoor Mobile Radio
  Commun. (PIMRC)}, Toronto, ON, Canada, 2023.

\bibitem{wenyan}
W.~Ma, L.~Zhu, and R.~Zhang, ``Passive beamforming for {3-D} coverage in
  {IRS}-assisted communications,'' \emph{IEEE Wireless Commun. Lett.}, vol.~11,
  no.~8, pp. 1763--1767, Aug. 2022.

\bibitem{outage}
T.~Van~Chien, A.~K. Papazafeiropoulos, L.~T. Tu, R.~Chopra, S.~Chatzinotas, and
  B.~Ottersten, ``Outage probability analysis of {IRS}-assisted systems under
  spatially correlated channels,'' \emph{IEEE Wireless Commun. Lett.}, vol.~10,
  no.~8, pp. 1815--1819, Aug. 2021.

\bibitem{weimultiirs}
W.~Mei and R.~Zhang, ``Joint base station and {IRS} deployment for enhancing
  network coverage: A graph-based modeling and optimization approach,''
  \emph{IEEE Trans. Wireless Commun.}, vol.~22, no.~11, pp. 8200--8213, Nov.
  2023.

\bibitem{10439018}
M.~Fu, W.~Mei, and R.~Zhang, ``Multi-passive/active-{IRS} enhanced wireless
  coverage: Deployment optimization and cost-performance trade-off,''
  \emph{IEEE Trans. Wireless Commun.}, \emph{early access}, 2024.

\bibitem{WdIRS}
Q.~Wu and R.~Zhang, ``Beamforming optimization for wireless network aided by
  intelligent reflecting surface with discrete phase shifts,'' \emph{IEEE
  Trans. Commun.}, vol.~68, no.~3, pp. 1838--1851, Mar. 2020.

\bibitem{OptYG}
G.~Yan, L.~Zhu, and R.~Zhang, ``Passive reflection optimization for {IRS-}aided
  multicast beamforming with discrete phase shifts,'' \emph{IEEE Wireless
  Commun. Lett.}, vol.~12, no.~8, pp. 1424--1428, Aug. 2023.

\bibitem{3GPPTS38}
``{NR}; physical layer measurements (release 18),'' \emph{Standard
  3GPP-TS-38.215, V18.0.0}, Sep. 2023.

\bibitem{Tobler}
W.~R. Tobler, ``A computer movie simulating urban growth in the {Detroit}
  region,'' \emph{Economic geography}, vol.~46, no. sup1, pp. 234--240, Jun.
  1970.

\bibitem{Kriging}
N.~Cressie, ``The origins of {Kriging},'' \emph{Mathematical geology}, vol.~22,
  pp. 239--252, Apr. 1990.

\bibitem{worboysgis}
M.~F. Worboys and M.~Duckham, \emph{GIS: a computing perspective}.\hskip 1em
  plus 0.5em minus 0.4em\relax Oxon, U.K.: CRC Press, 2004.

\bibitem{chi1999}
J.~Chiles and P.~Delfiner, \emph{Geostatistics: Modelling spatial
  uncertainty}.\hskip 1em plus 0.5em minus 0.4em\relax New York, U.S.: Wiley
  Press, 1999.

\bibitem{rumelhart1986learning}
D.~E. Rumelhart, G.~E. Hinton, and R.~J. Williams, ``Learning representations
  by back-propagating errors,'' \emph{Nature}, vol. 323, pp. 533--536, Oct.
  1986.

\bibitem{haykinneural}
S.~S. Haykin, \emph{Neural networks and learning machines}, 3rd~ed.\hskip 1em
  plus 0.5em minus 0.4em\relax New Jersey, U.S.: Pearson Education Press, 2009.

\bibitem{tao2020intelligent}
Q.~Tao, S.~Zhang, C.~Zhong, and R.~Zhang, ``Intelligent reflecting surface
  aided multicasting with random passive beamforming,'' \emph{IEEE Wireless
  Commun. Lett.}, vol.~10, no.~1, pp. 92--96, Jan. 2021.

\end{thebibliography}

\end{document}